\documentclass[fleqn,10pt]{wlscirep}
\usepackage[utf8]{inputenc}
\usepackage[T1]{fontenc}
\usepackage{amsmath}
\usepackage{array}
\usepackage{amssymb}
\usepackage[tight,footnotesize]{subfigure}
\usepackage{amsthm}
\usepackage{tabularx}
\usepackage{graphicx}
\usepackage{graphics}
\usepackage{subfigure}
\usepackage{float}
\usepackage{cite}
\usepackage{framed}
\usepackage{authblk}

\title{Distress propagation on production networks: Coarse-graining and modularity of linkages}

\author[1]{Ashish Kumar}
\author[2,*]{Anindya S. Chakrabarti}
\author[3,**]{Anirban Chakraborti}
\author[4]{Tushar Nandi}
\affil[1]{Economics Department, Ashoka University, Sonipat-131029, Haryana, India}
\affil[2]{Economics Area, Indian Institute of Management, Ahmedabad-380015, Gujarat, India}
\affil[3]{School of Computational and Integrative Sciences, Jawaharlal Nehru University, New Delhi-110067, India}
\affil[4]{Economics Department, Indian Institute of Science Education and Research Kolkata, Mohanpur-741246, West Bengal, India}
\affil[*]{anindyac@iima.ac.in }
\affil[**]{anirban@jnu.ac.in}

\begin{abstract}
\noindent 
Distress propagation occurs in connected networks, its rate and extent being dependent on network topology. To study this, we choose economic production networks as a paradigm. An economic network can be examined at many levels-- linkages among individual agents (microscopic), among firms/sectors (mesoscopic) or among countries (macroscopic). New emergent dynamical properties appear at every level, so the granularity matters. For viral epidemics, even an individual node may act as an epicenter of distress and potentially affect the entire network. Economic networks, however, are known to be immune at the micro-levels and more prone to failure in the meso/macro-levels. We propose a dynamical interaction model to characterize the mechanism of distress propagation, across different modules of a network, initiated at different epicenters. Vulnerable modules often lead to large degrees of destabilization. We demonstrate our methodology using a unique empirical data-set of input-output linkages across 0.14 million firms in one administrative state of India, a developing economy. The network has multiple hub-and-spoke structures that exhibits moderate disassortativity, which varies with the level of coarse-graining. The novelty lies in characterizing the production network at different levels of granularity or modularity, and finding `too-big-to-fail' modules supersede `too-central-to-fail' modules in distress propagation. 

\end{abstract}
\begin{document}

\renewcommand{\theenumi}{(\roman{enumi})}%

\flushbottom
\maketitle

\thispagestyle{empty}

%\noindent Please note: Abbreviations should be introduced at the first mention in the main text – no abbreviations lists. Suggested structure of main text (not enforced) is provided below.

%%%%%%%%%
%\input{paper_main}
%%%%%%%%%
\section*{Introduction}

\begin{quote}
	\textit{``The current crisis is more complicated. It is not a financial crisis, but a real crisis affecting the supply and demand system. $\ldots$ If the supply chain is broken, an increase in demand will not solve the problem.''} 
	\newline \hspace*{0pt}\hfill -- Joseph Stiglitz on coronavirus pandemic and its aftermath\cite{Stiglitz_2020} [17th March, 2020]
\end{quote}

The coronavirus (COVID-19) pandemic has brought the entire world to a standstill; a crisis of gigantic proportion and unprecedented. This could happen at a rapid pace and to such a great extent because of the extremely dense global connectivity in the human socio-technological world. A complex interdependent system is often `robust yet fragile'\cite{lux2009economics} and its fragility with respect to external or internal perturbations may represent emergent phenomena.\cite{arthur1999complexity,chialvo2010emergent}
Both robustness and fragility of a complex system can originate from the network of interdependencies among the constituents and,
the mechanism of distress propagation in turn, is known to be dependent on its topological properties, ranging from the node-level local characteristics to the network-level global characteristics. \cite{starnini2019interconnected,mones2014shock,bardoscia2016distress}
On one extreme, the failure of a \textit{single individual} node might cause a cascading failure in the whole network, e.g., power grid, traffic system, societal spreading of rumor/infection.\cite{wang2019local,mones2014shock,mamede2012overspill,rudenberg1968electrical,newman2002spread}
At the other extreme of systemic robustness with respect to localized external perturbations, we have economic networks, where the fluctuations and economic cycles of boom and bust are thought to be driven by \textit{aggregate} shocks like monetary shocks, productivity fluctuations and so forth.\cite{mankiw1989real}
However, this traditional view of an economic crisis has been challenged in the aftermath of the recent financial downturn of 2007-09. In an important paper, Gabaix\cite{Gabaix_2011_Granular} showed theoretically and empirically that aggregate fluctuations in an economy could be driven by firm-level shocks. A complementary approach was proposed by Acemoglu et al.\cite{Acemoglu_Network_2012} where they showed that the phenomenon of shock propagation through sectoral linkages 
could impact aggregate volatility. Both of the seminal works put forward
the role of {\it granular} economic entities in terms of explaining systemic instability.

At the onset, it may be useful to unravel the differences between ``aggregate'' shocks and ``idiosyncratic'' shocks in an economy.
The idea of aggregate shocks\cite{lucas1977understanding} is quite appealing: if production clusters (say, firms or sectors) have uncorrelated disturbances, such disturbances would tend to average out as we disaggregate the economy into finer and finer levels of production, essentially due to the {\it central limit theorem}. This in turn, would imply that only aggregate shocks can have macroeconomic consequences. Indeed, the source of fluctuations in most of the business cycle literature can be traced back to aggregate supply side or demand side shocks.\cite{horvath2000sectoral,acemoglu2016networks,horvath1998cyclicality} 
In contrast, there is new literature that emphasizes the inter-linkages in the production structure of modern economies and how they might actually amplify rather than mute the idiosyncratic effects arising out of production units.\cite{acemoglu2016networks} 

In this paper, we examine the role of network architecture (modularity) of a granular economy on the origin of a shock and how it percolates in an interdependent economy, which can be considered as an ideal example of a complex system. We combine features from both the approaches and show that clusters of firms (modules), which are closely connected through supply chains, can serve as a  model for understanding the effects of shock or distress propagation. 
In particular, we study the architectures of the firm-to-firm production network at different levels of coarse-grained filters and the resultant shock propagation mechanisms. This method can then be easily generalized  and extended to the study of distress propagation in other complex systems, and hence may be considered as a prototype.

Our main proposition is that the phenomena of distress propagation is fundamentally dependent on the mesoscopic properties
of the production network. In particular, the modular architecture of the network should be seen as the basic building block. Below we explain why and how it differs from the firm-level\cite{Gabaix_2011_Granular} and sector-level\cite{Acemoglu_Network_2012} granularity. First of all, we note that potential {\it shocks} might be local in nature affecting individual firms. A {\it sector} is a collection of firms that produces similar goods classified under the same type of production processes, which is more coarse-grained and granular than a firm.     
Firms on the other hand, might be made of production plants. Ideally, we should be able to trace the epicenter of a shock to the plant-level, which is literally the smallest production unit. However, such fine-grained data is extremely rare and largely not available. Here, we have utilized an unique administrative data-set obtained from the Indian state of West Bengal, which was generated by matching the tax-collection records of nearly 0.14 million firms while they carried out bilateral trades of purchasing and selling of manufactured goods. We have constructed the networks from the data collected in four quarters of the fiscal year 2016. The time-period considered is special as on 8th November 2016, the Indian government carried out {\it demonetization}, which entailed the existing currencies to be declared ineffective and not backed by the government. The goal was to reduce the extent of the shadow economy and the use of counterfeit cash. Since the Indian economy does not yet have a well-developed financial market and trades mostly depend on cash transactions, such a move can be expected to have an effect on bilateral trades between production units.

A summary of our main results is as follows: 
\begin{enumerate}
\item Macroscopic properties of the production network are quite stable. The demonetization shock did not have an impact on the network as a whole, at least immediately after the shock. Therefore, the network exhibits {\it stickiness}, which is economically significant.
\item There is considerable heterogeneity in the local topological characteristics of the firms' connectivities in the
network. The nature of dispersion in the number of the first-order and second-order neighbors closely mimic the same found in sectoral data.\cite{acemoglu2016networks} 
\item The empirical production network is modular. We construct networks of modules by coarse-graining, where each module comprises a set of firms tightly interlinked through supply chains.
Based on the modular network of communities, we show using simulations of distress propagation that `too-big-to fail' modules are more prone to cascading failures than `too-central-to fail' modules. Though counter-intuitive, this is significant for understanding the mechanics of distress propagation in complex networks.
\end{enumerate}

Below we elaborate the significance of our results and place them in the context of the emerging field in the intersection of network science and economic fluctuations.
\begin{enumerate}
\item \textbf{Complex network science}

\begin{itemize}
\item 
Firstly, there is a large literature on theoretical models of network formation that mimic real-world networks.\cite{newman2006structure}
In our case, a particularly important class of models are ``scale-free networks''\cite{barabasi2016network}, as the firm-to-firm network shows power law-like behavior in the degree distribution (both in first order and second order). The entire production network shows mild disassortativity.
However, once we construct the network of communities from the production linkages, then the resulting network exhibits disassortativity. Therefore, it indicates the existence of multiple hubs indirectly connected to each other. We show that this finding has important implications for the mechanism of distress propagation.
\item
Secondly, there is a growing literature on statistical description of empirical networks constructed from economic and financial data.\cite{sharma2017financial,kumar2020disentangling,abergel2012econophysics} Textbook treatments are also available, although mostly in the context of financial data.\cite{sinha2010econophysics,newman2006structure} Our paper  fits well in this literature, as it provides an
unique network view of the production process in developing economies. As far as we know, there is barely any documentation of firm-to-firm network available in the context of developing economies. This has become even more important in the context of the present WHO-declared pandemic of coronavirus (COVID-19). As we quoted above the Nobel laureate Joseph Stiglitz, this pandemic will cause both demand side and supply side shocks\cite{Stiglitz_2020}. This will have serious repercussions.  The present study will be important as the literature on developing countries supply-chain architecture is not well-developed.
\end{itemize}

\item \textbf{Aggregate fluctuations in granular economy}

\begin{itemize}

\item Firstly, our results relate to the literature on the granular architecture of economies, which started with a theoretical debate on whether granularity and sectoral inter-linkages matter for aggregate dynamic properties.\cite{horvath1998cyclicality,horvath2000sectoral,foerster2011sectoral}
The work of Gabaix\cite{Gabaix_2011_Granular} brought the role of granularity of firms in explaining business cycle fluctuations at the forefront.
Foerster et al.\cite{foerster2011sectoral} 
modeled an economy with standard business cycle model with an explicit input-output network,
which was mapped directly to the industrial production and matched the model results with empirical decomposition of aggregate shocks and idiosyncratic sector specific shocks, showing evidence in favor of granularity. 
\item Secondly, there is a stream of work on utilizing observable large shocks on a set of firms or industries and tracing their impact through the input-output network; e.g., David\cite{david2013china} studied the impact of increased Chinese competition into the U.S. economy through input-output linkages and local labor markets. Barrot\cite{barrot2016input} and Carvalho\cite{carvalho2016supply} focus on the transmission of shocks arising out of natural disasters, such as 2011 Japanese earthquake, over the global input-output network.

\item 
Thirdly, we reiterate that the main message of the paper is the proposed link between modularity of networks and the mechanism of distress propagation, which is a novel way of coarsening production units and studying distress propagation, as opposed to more standard ideas of coarsening at the level sectors.\cite{acemoglu2016networks,foerster2011sectoral,horvath1998cyclicality,horvath2000sectoral} This idea resonates with the literature of propagation of location-specific shocks (e.g. natural disasters\cite{carvalho2016supply}) through the supply-chain network.
However, exact causal identification of shock propagation could not be done in the present context due to limited availability of data, and the fact that the only identifiable shock of {\it demonetisation} was 
an \textit{aggregate} shock (separation of control and treatment group in the traditional sense of natural experimentation was not possible). Still the simulation results along with the detailed topological description provide a fairly complete picture of the production network and its impact of volatility or fluctuations.
\end{itemize}
\end{enumerate}

\section*{Methodology and Results}
\subsection*{Construction of the production network}
There are about 0.14 million firms in our data universe, with about half a million trade linkages among themselves, in each of the quarters (see Data description in Methods). In constructing the weighted network, firms are {\it nodes} and the trade linkages between a pair of firms represent an {\it edge} between the corresponding pair of nodes, with the edge weight being the normalized transaction volume.

\begin{table}[]
	\centering
	\caption{Characteristics of production network for the Indian state of West Bengal for quarters 1-4 in the fiscal year 2016}
	 	\resizebox{0.8\linewidth}{!}{%
	\begin{tabular}{clllllll} 
		\toprule
		quarter & nodes   & edges  & \begin{tabular}[c]{@{}l@{}}average \\degree \end{tabular}  & \begin{tabular}[c]{@{}l@{}}number \\of clusters\end{tabular} & \begin{tabular}[c]{@{}l@{}}nodes in the giant \\component (GC) \end{tabular} & \begin{tabular}[c]{@{}l@{}}edges \\(GC)\end{tabular} & \begin{tabular}[c]{@{}l@{}}average \\degree (GC)\end{tabular} \\ 
		\hline
		1       & 132331 & 510746 & 7.72 & 483                                                   & 131148                                                               & 510026 & 7.78                                                                                                                                                              \\ 
		\hline
		2       & 132537 & 510178 & 7.61 & 500                                                   & 131303                                                               & 509404 & 7.67                                                                                                                                                              \\ 
		\hline
		3       & 138660 & 541960 & 7.725 & 503                                                   & 137482                                                                & 541254 & 7.781                                                                                                                                                             \\
		\hline
		4       & 144509 & 582651 & 8.06 & 424                                                   & 143518                                                                & 582045 & 8.11                                                                                                                                                             \\
		\bottomrule
	\end{tabular}
}
   \label{table:summary_stats}
\end{table}

We provide the summary statistics for the entire production network in Table \ref{table:summary_stats}.
Columns 6-8 in Table \ref{table:summary_stats} show that giant component or maximally connected component comprises around 99\% of the production entities along with almost all trade linkages that were present in the original network. Note that, by construction, every entity in the giant component has a path connected to every other production entity in the network. Theoretically, 
distress originating in any part of a sub-graph of the original network can propagate to any other sub-graph. The remaining 1\% of the peripheral nodes cannot participate in distress propagation, and hence were discarded from our analyses.
It may be mentioned here that while extracting the giant component from the original network, the edges were treated as \textit{undirected}, such that the trade linkages possess information on both directions: input flows (from seller to buyer) and monetary flows (from buyer to seller). We found that giant components are quite stable in terms of aggregate properties, across the time periods considered.

\subsection*{Topology of the coarse-grained structures}
The production network is a huge set of 0.14 million nodes and 0.5 million edges. In order to analyze the granular nature of the network, we have conducted our analyses at two levels of coarse-grained filtering: 
\begin{enumerate}
\item
We have extracted the {\it backbone} of the giant component using {\it disparity filter}\cite{serrano2009extracting}, which essentially exploits local heterogeneity and local correlations among the edge weights across nodes and preserves a subset of edges that survive at all the scales of the parent network (see Methods). There is a tuning parameter $\alpha$ that can be set as an input, in order to control the number of nodes and edges in the filtered backbone. There is an arbitrariness in the choice of $\alpha$, depending on the scale of coarse-graining we would like to obtain. The choice of this cut-off parameter  may be further tuned, if we have sectoral information and other details  of the production firms (which we do not have presently).
For our analyses, we have kept $\alpha$ = 0.0001 and we have checked that our results are robust as far as small changes in $\alpha$.  

\item We have employed community detection algorithms to extract the meso-level properties of the network. In particular, we employ {\it modularity maximization} and {\it Infomap} algorithms, which are more coarse-grained than the backbone and reduces the dimension of the original network substantially. We construct the {\it networks of communities} through two different algorithms: modularity maximization and Infomap algorithms (see Methods).
\end{enumerate}

\begin{figure}[]
 	\centering
\includegraphics[width=0.74\linewidth]{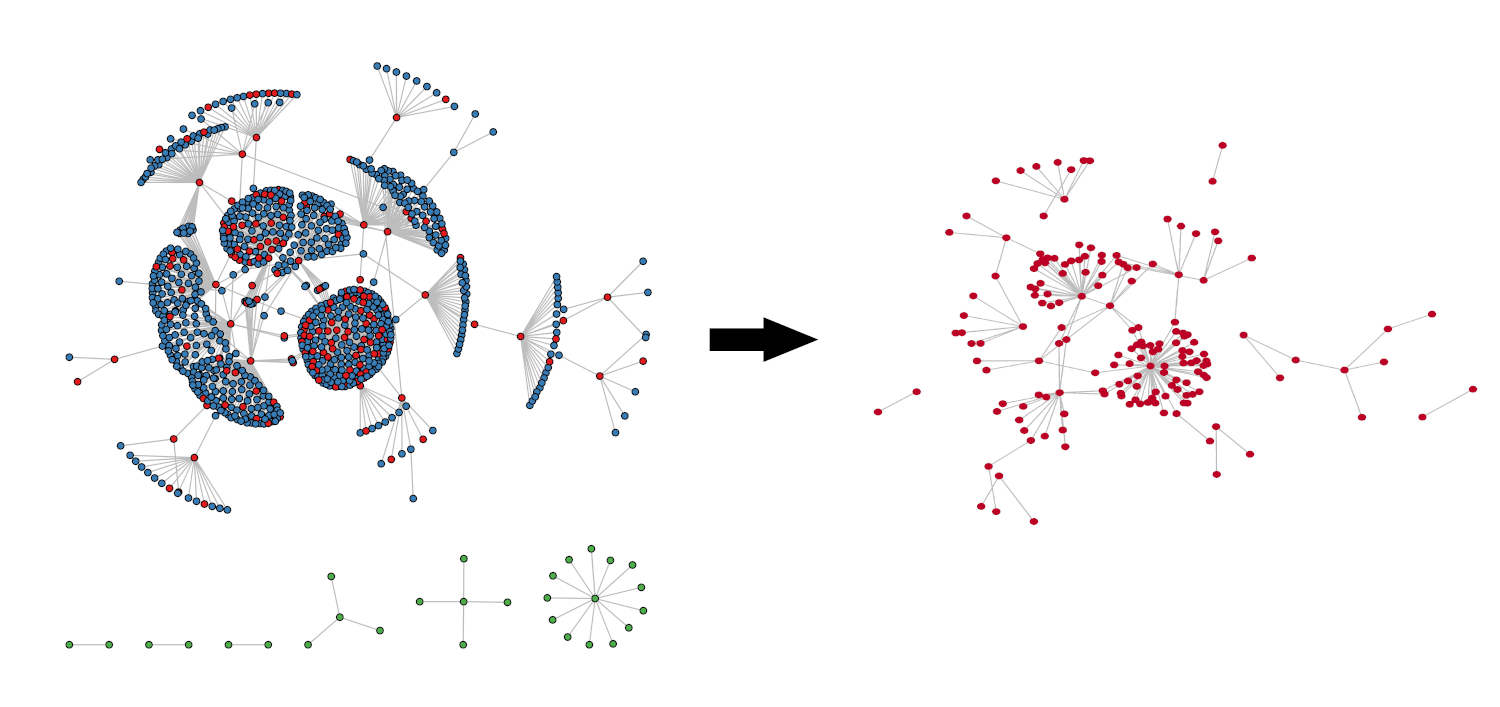}\llap{\parbox[b]{5.4in}{{\large \textsf{a}}\\\rule{0ex}{2in}}}\\
\includegraphics[width=0.74\linewidth]{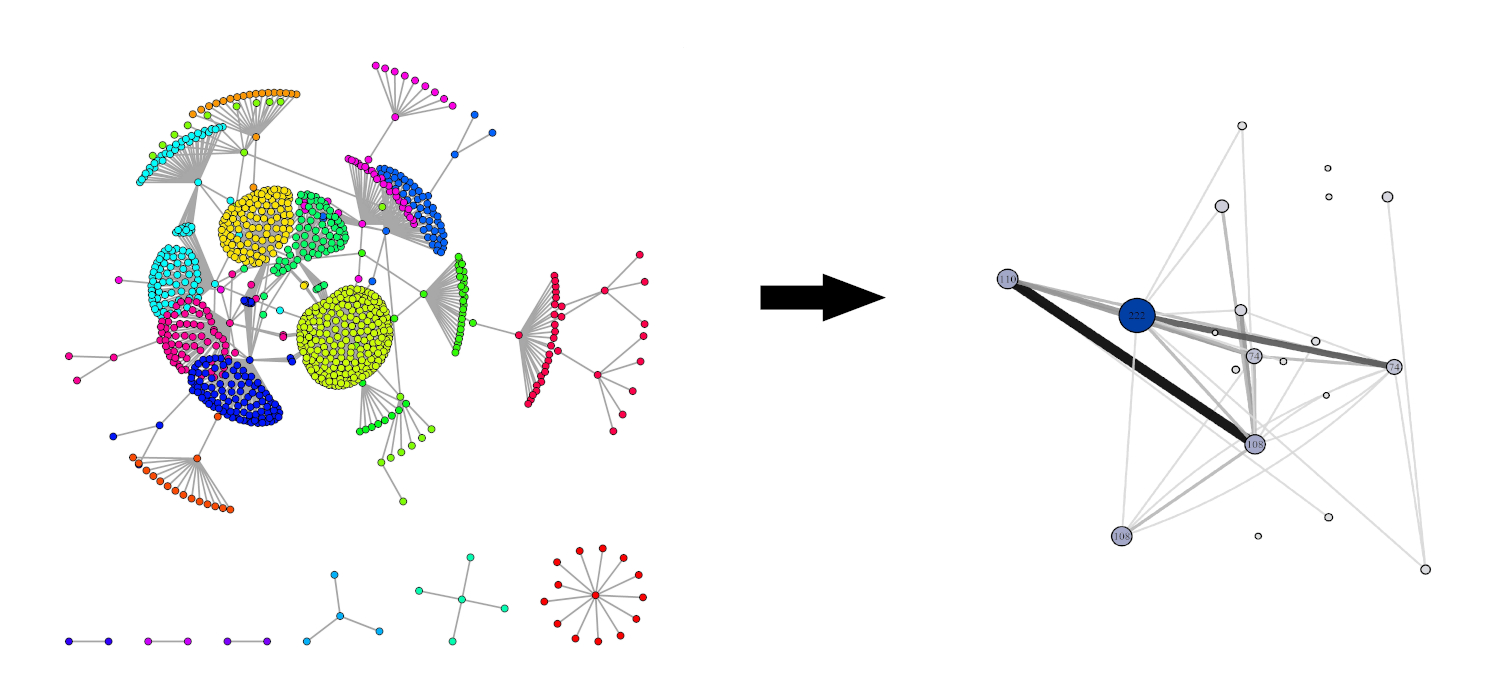}\llap{\parbox[b]{5.4in}{{\large \textsf{b}}\\\rule{0ex}{2in}}}\\
\includegraphics[width=0.74\linewidth]{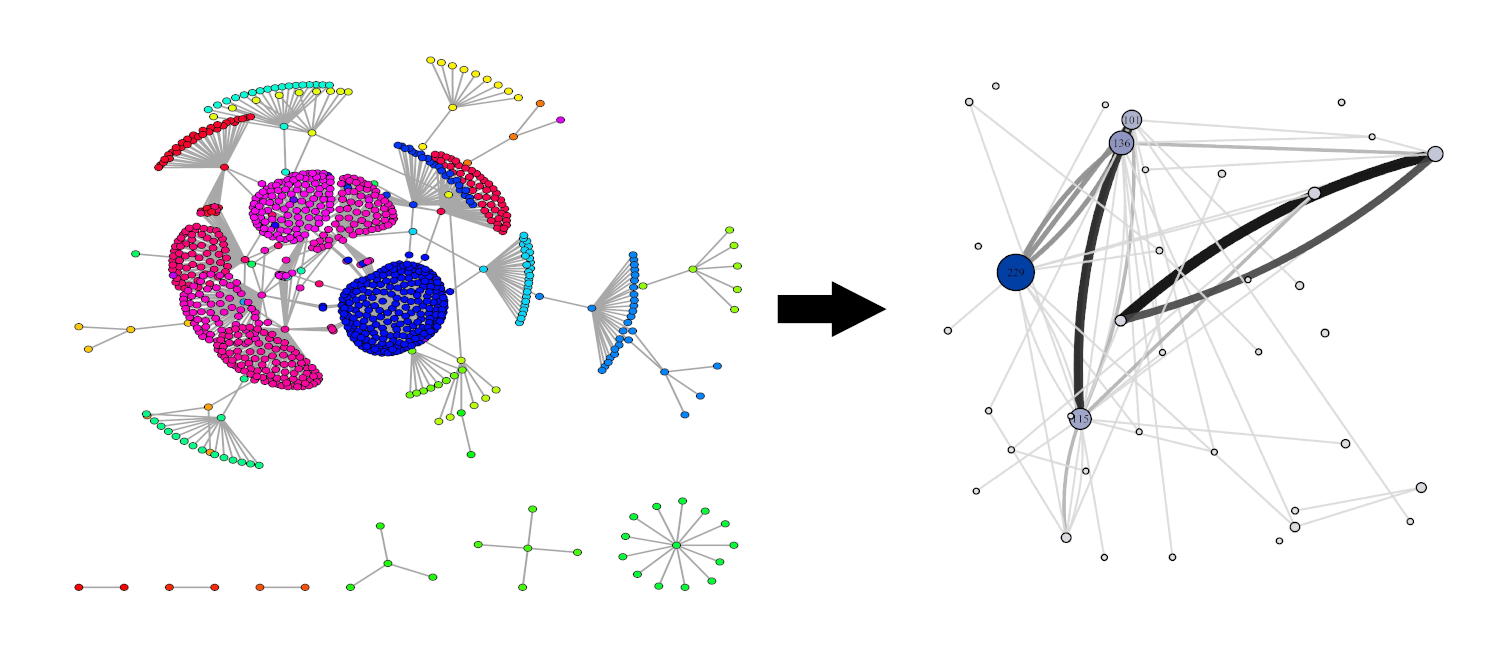}\llap{\parbox[b]{5.4in}{{\large \textsf{c}}\\\rule{0ex}{2in}}}
			\caption{\textbf{Schematic diagram to show Coarse-grained networks: (a) Extraction of the backbone from the production network, (b) Community detection using modularity maximization algorithm and the resultant network of communities, and (c) Community detection using Infomap algorithm and the resultant network of communities}.
				Panel (a): The figure shows a sample network obtained from the data (954 nodes and 1000 edges). The network has three components -- (i) giant component colored blue (925 nodes and 977 connections), (ii) unconnected peripheral nodes colored green (29 nodes with 23 edges), and (iii) the backbone (red) extracted using disparity filter\cite{serrano2009extracting} (171 nodes and 155 edges). 
				In panels (b) and (c), we shows the {\it network of communities} extracted from the production network.
				Each community is comprises of a set of firms and is collapsed into one node in the {\it network of communities}. The size of each node is scaled according to the number of firms in it. The edges are weighted according to the number of connections between the constituent firms belonging to these communities.
				Panel (b): 20 communities are extracted using modularity maximization algorithm. The largest community has 222 firms in it. Panel (c): 38 communities extracted using Infomap algorithm. The largest community has 229 firms in it.}
   \label{fig:combined_backbone_clustering}
\end{figure}

The scheme of steps involving transformation of the entire production network to its granular representations at various levels of coarseness, is exhibited in Fig.~\ref{fig:combined_backbone_clustering}. 
To give a visual idea of the resulting networks, we have sampled a small sized network consisting of 954 nodes (firms) with 1000 edges (trade linkages) among them, from the original data. Panel (a) shows extraction  of the backbone via disparity filter.\cite{serrano2009extracting} Panels (b) and (c) exhibit the {\it network of communities} constructed using 
modularity maximization and Infomap algorithms, respectively.

\subsection*{Topological properties of the input-output structures}
The topological structure of the network is seen to be very persistent across the four quarters (see Table \ref{table:summary_stats})  despite facing a massive exogenous economic shock in the form of demonetization. Although the number of firms and trade linkages have surged around the third quarter to a certain extent but overall the network is quite stable in terms of other clustering properties along with localized degree connectivity.

 \begin{table}[]
 	\centering
 	\caption{Characteristics of the backbone ($\alpha$ =  0.0001) extracted from GC of the production network for the Indian state of West Bengal for quarters 1-4 in the fiscal year 2016}
 	\resizebox{0.68\linewidth}{!}{%
 		\begin{tabular}{cllllll} 
 			\toprule
 			quarter & nodes (\%)   & edges (\%) & \begin{tabular}[c]{@{}l@{}}number of \\clusters (\%)\end{tabular} & nodes in GC & \begin{tabular}[c]{@{}l@{}}average \\degree \end{tabular} & \begin{tabular}[c]{@{}l@{}} normalized\\ weights  \end{tabular}  \\ 
 			\hline
 			1       & \begin{tabular}[c]{@{}l@{}}19032~\\(14.4)\end{tabular}  & \begin{tabular}[c]{@{}l@{}}14993\\(2.9)\end{tabular}  &   4788                                                       &  6488                                                              & 2.22                                                                   & 0.47                                                                                       \\ 
 			\hline
 			2       & \begin{tabular}[c]{@{}l@{}}27126~\\(20.4)\end{tabular}  & \begin{tabular}[c]{@{}l@{}}23920\\(4.7)\end{tabular}  &   5291                                                        &  13446                                                               & 1.759                                                                   & 0.38                                                                                       \\ 
 			\hline
 			3       & \begin{tabular}[c]{@{}l@{}}29811~\\(21.4)\end{tabular} & \begin{tabular}[c]{@{}l@{}}25862\\(4.8)\end{tabular}  & 6088                                                           & 13828                                                                & 
 			1.731                                                                    & 0.39                                                                                       \\
 			\hline
 			4       & \begin{tabular}[c]{@{}l@{}}21920~\\(15.6)\end{tabular} & \begin{tabular}[c]{@{}l@{}}17492\\(3)\end{tabular}  & 5411                                                          & 7915                                                                & 
 			2.24                                                                    & 0.50                                                                                       \\
 			\bottomrule			
 		\end{tabular}
 	}
  \label{table:summary_stats_BB}
 \end{table}

Table \ref{table:summary_stats_BB} shows basic topological features of the backbone network extracted from the giant component in quarters 1-4. We note that 2.9-4.8\% of the trade linkages (column 3) in backbone carry 39-50\% of the transactional value of the all the trade linkages (column 7) present in the entire production network. Additionally, only 14.4-21.4\% of the production firms feature in the backbone in comparison to the entire network.

\subsubsection*{First-order degree distributions}

\begin{figure}[]
	 			\centering
 			\subfigure[Firm-to-firm network: In-Degree]{
 				\includegraphics[width=0.3\textwidth]{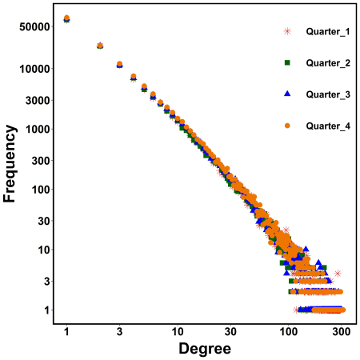}
 			}
 		\subfigure[Giant component: In-Degree]{
 			\includegraphics[width=0.3\textwidth]{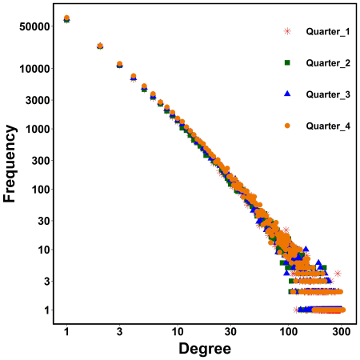}
 		}
 	     \subfigure[Backbone: In-Degree]{
 		\includegraphics[width=0.3\textwidth]{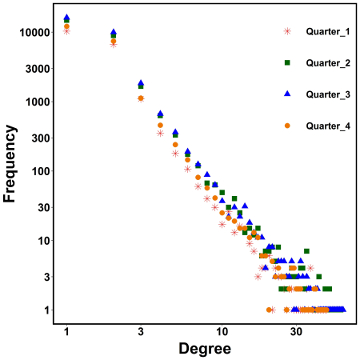}
        	}
        \subfigure[Firm-to-firm network: Out-Degree]{
        	\includegraphics[width=0.3\textwidth]{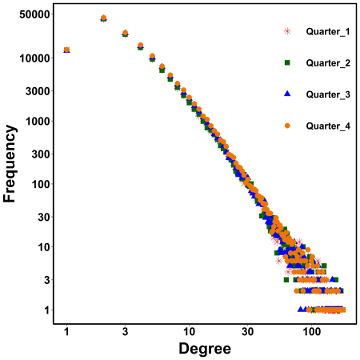}
        }
        \subfigure[Giant component: Out-Degree]{
        	\includegraphics[width=0.3\textwidth]{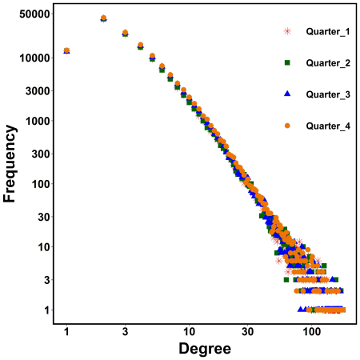}
        }
        \subfigure[Backbone:Out-Degree]{
        	\includegraphics[width=0.3\textwidth]{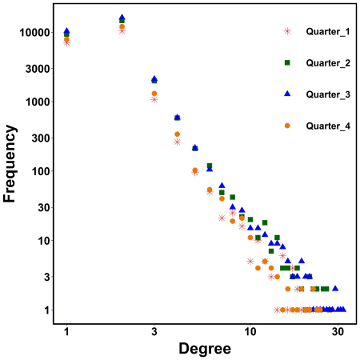}
        }
 	\caption{\textbf{In-degree and out-degree distributions}. Plots in log-log scale for the four quarters in the fiscal year 2016; before and after the demonetization: (a) and (d) Entire production network,
 		(b) and (e) Giant component,
 		(c) and (f) Backbone. In all three cases, the distributions are quite stable even after the shock of {\it demonetisation} between the second and third quarters. Out-degree seems to have exhibit a slightly steeper decay than in-degree in all cases, indicating lesser degree of heterogeneity.}
     \label{fig:Degree_Dist}
\end{figure}

In Figure.~\ref{fig:Degree_Dist}, we show 
the degree distributions for the entire production network, the giant component (GC) and the backbone across both quarters. The $x$-axis denotes log of degree (in and out) and the $y$-axis denotes the 
log frequency.

As the empirical distribution suggests, there are three main features in the data: (i) There is substantial heterogeneity across the firms in terms of degree distribution. The almost linear nature of the frequency plot is indicative of a fat-tailed distribution. Therefore, some firms are disproportionately more important than other firms in the buyer-seller network. (ii) In-degree and the out-degree seem to be distributionally similar to each other except that in all cases, the out-degree has a mildly sharper decay. (iii) The phenomenon of demonetisation seems to have had no impact on the degree distributions. The distributions in quarters 2 and 4 have extensive overlaps.

\subsubsection*{Second-order degree distribution}
 
For second-order degree distributions, we refer to Fig.~S1 in the supplementary material, where we have presented the results for the backbone network only.
Similar to the first-order degree distributions, second-order degree distribution also have relatively heavy tails in all the quarters. It is noteworthy in this context that
Acemoglu et al. (2012)\cite{Acemoglu_Network_2012}
showed that the fat tail in the second-order connectivity contributes to aggregate volatility. 
In their case, they had shown it in the context of sectoral data whereas our results utilize
firm-level trade linkages. 

Given the fat tailed nature of the first-order and second-order degree distributions in the original network as well as the filtered networks, we interpret the results to indicate that the topology of the network would play major role in distress propagation across the network.
Disruption in one firm, who is not only supplier to a large number of firms, but also is supplier to other firms who are large suppliers, would potentially propagate from the epicenter to a large cluster of firms within network distance 2.

\subsubsection*{Degree distribution in networks of communities}

Fig.~S2 exhibits the degree distribution for the network of communities extracted using both modularity maximization and Infomap algorithms from the backbone of production network for two quarters. 
The degree distributions corresponding to the modularity maximization algorithm has much narrower range of variation than the one extracted using Infomap algorithm.
This difference can be attributed to the resolution limit of the modularity maximization algorithm 
leading to smaller clusters.
Therefore, Infomap provides a more coarse-grained clustering structure than the modularity maximization algorithm.

\begin{table}[]
	\centering
	\caption{Assortativity analysis for the coarse-grained networks}
%	\resizebox{\linewidth}{!}{%
		\begin{tabular}{>{\hspace{0pt}}p{0.10\linewidth}>{\hspace{0pt}}p{0.10\linewidth}>{\hspace{0pt}}p{0.10\linewidth}>{\hspace{0pt}}p{0.10\linewidth}>{\hspace{0pt}}p{0.15\linewidth}>{\hspace{0pt}}p{0.15\linewidth}} 
			\toprule
			quarter & production\par{}network & GC & backbone & com. network \par{}(mod. max.) & com. network \par{}(Infomap)  \\ 
			\hline
			1       & -0.067                               &-0.067               & -0.19 & -0.12 &  0.013 \\ 
			\hline
			2       & -0.07                               & -0.07                & -0.13 & -0.21 &  0.015 \\ 
			\hline
			3       & -0.063                              & -0.063                 & -0.14 & -0.15 & 0.031  \\
			\hline
			4       & -0.062                              &  -0.062                & -0.16 & -0.15 & 0.08  \\
			\hline
		\end{tabular}
%	}
	\label{table:assortativity}
\end{table}

In Table \ref{table:assortativity}, we report the assortativity coefficients of the networks at different levels of granularity. We see that the entire network along with the giant component, are mildly disassortative. However, the backbone network is quite strongly disassortative. the community networks obtained through modularity maximization are strongly disassortative whereas the community network obtained from Infomap algorithm is very mildly assortative in nature.
The disassortative nature of the network of communities (along with large heterogeneity in degree distribution) is important in the context of shock propagation, since it indicates that there are multiple hub-and-spoke structure embedded in the network where the hubs are indirectly connected. Therefore, shocks generated in one hub can transmit to other hubs and the corresponding spokes. Below, we formally address the phenomena of shock propagation in the context of a dynamical model.

\subsection*{Modularity and shock propagation: Simulation results}

\subsubsection*{Vector autoregression model}
In order to model shock propagation in an interlinked economy, we rely on a simple statistical dynamic model called vector autoregression model (VAR).\cite{hamilton1994time} 
This model allows us to calibrate the inter-linkages to the observed empirical network and thereby provide unique identification of paths for shock propagation from any given epicenter to the rest of the network. 

The $VAR(p)$ model is a theoretical representation of the evolution of $n$ interacting variables (denoted by $n\times 1$ column vector $x_{t}$) dependent on their lagged values of the order $1 \leq....\leq p$ and a vector of noise ($u_t$):
\begin{equation}
 x_{t} = A_{1}x_{t-1} +.... + A_{p}x_{t-p} + u_{t},
 \label{eqn:var_matrix}
\end{equation}
where $x_{t} = (x_{1t},...,x_{nt})$ is a column vector indexed by time period $t$, the coefficients of interaction terms are collected in square matrices $A_{k}$ where $k$ corresponds to the lag order, $u_{t} = (u_{1t},..,u_{nt})$ is a stationary noise vector ($\mathbf{E}(u_{t}) = \mu$ and covariance matrix $\mathbf{E}(u_{t}u_{t}^{T}) = \sum_{u}$).
 
The essential idea of application of VAR in the present context is that we can imagine that each firm has its own idiosyncratic shock process (fluctuations due to local demand and supply factors) and due to the interlinked nature of the shock process, one firm's shock gets transmitted to its first-order neighbors and then their first-order neighbors and so on. In order to implement the mechanism, we first impose a simple structure on the interlinkages:
\begin{equation}
x_{t} = A x_{t-1} + u_{t},
\end{equation}
where we have deliberately chosen a simple model with one lag ($p=1$ in Eqn. \ref{eqn:var_matrix}) to reduce dimensionality of the model and parameters. The matrix $A$ gives us the connectivity structure across the firms, which can be mapped to the empirical data.
 
In actual implementation, we calibrate the matrix $A$ to the adjacency matrix of the network of communities rather than the network of firms. The main reason for doing that is to reduce computational burden. Dealing with a system of equation with 0.14 million variables would be prohibitively expensive in terms of computation. The second reason is the whereas purely microeconomic shock sot individual firms may not have a substantial impact, a cluster-wide shock will surely have much more sizeable effect.
We utilize {\it impulse response functions} to measure the impact and propagation of a single idiosyncratic shock, across the network. We select one epicenter node and give a shock to it in a network through vector $u$. Then by iterating the VAR equation, we can capture the dynamic evolution of resulting propagation of the shock over time and across nodes.
In order to establish that different topology of the networks lead to different patterns of distress propagation,
we will first demonstrate the mechanism based on four toy models depicting rudimentary network structures.

\subsubsection*{Distress propagation in toy network structures}
\begin{figure}[]
 			\centering
 			\subfigure[Network Structure]{
 				\includegraphics[width=0.22\textwidth]{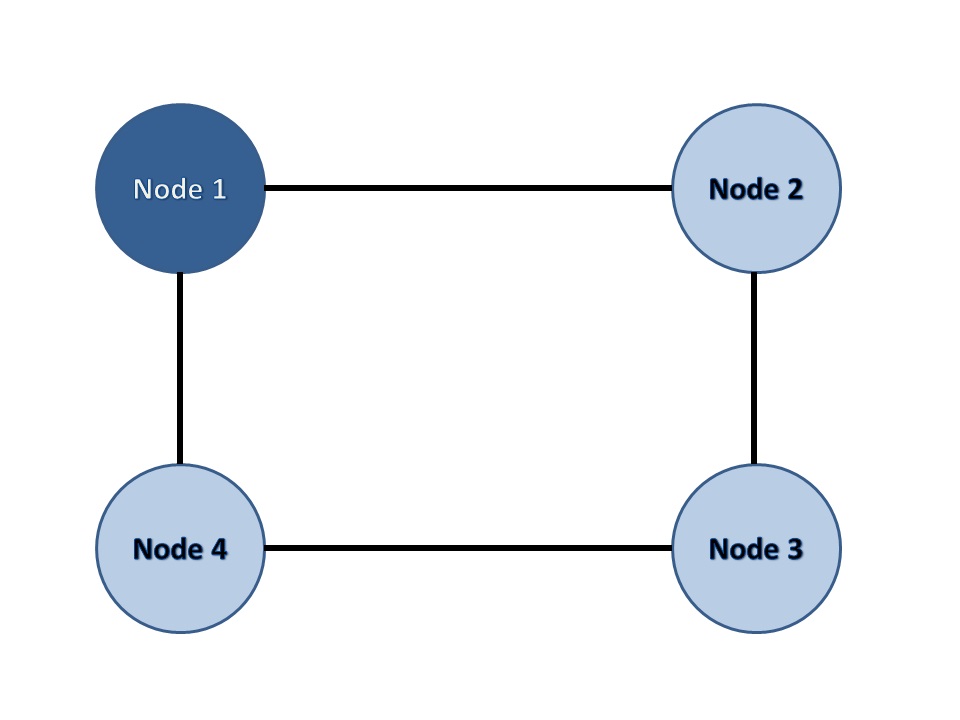}
 			} 
 			\subfigure[Network Structure]{
 				\includegraphics[width=0.22\textwidth]{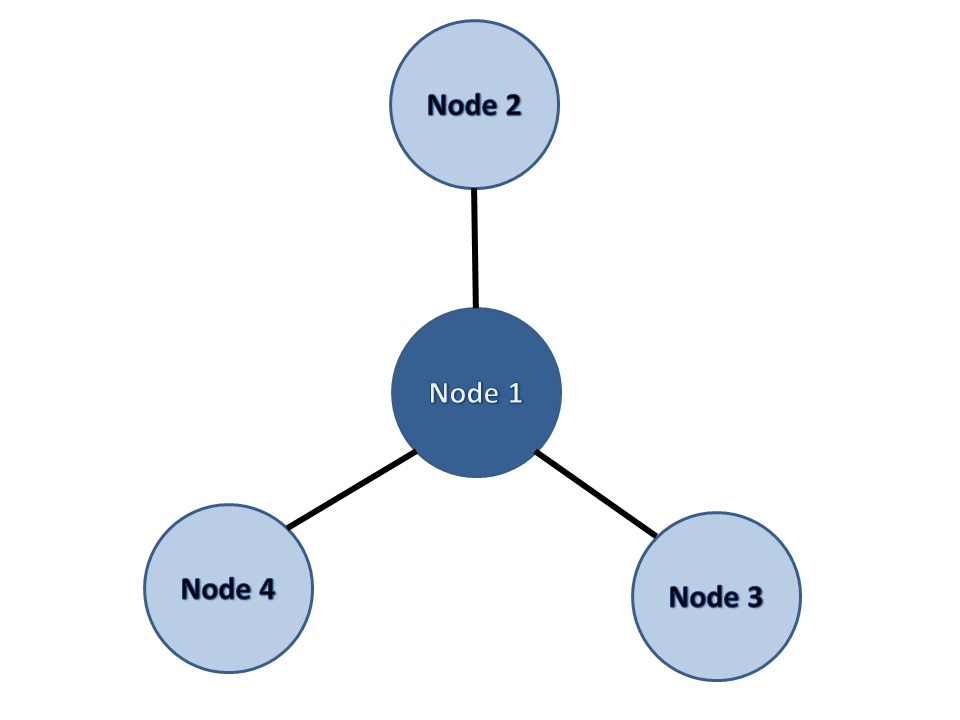}
 			} 
 		  \subfigure[Network Structure]{
 			\includegraphics[width=0.22\textwidth]{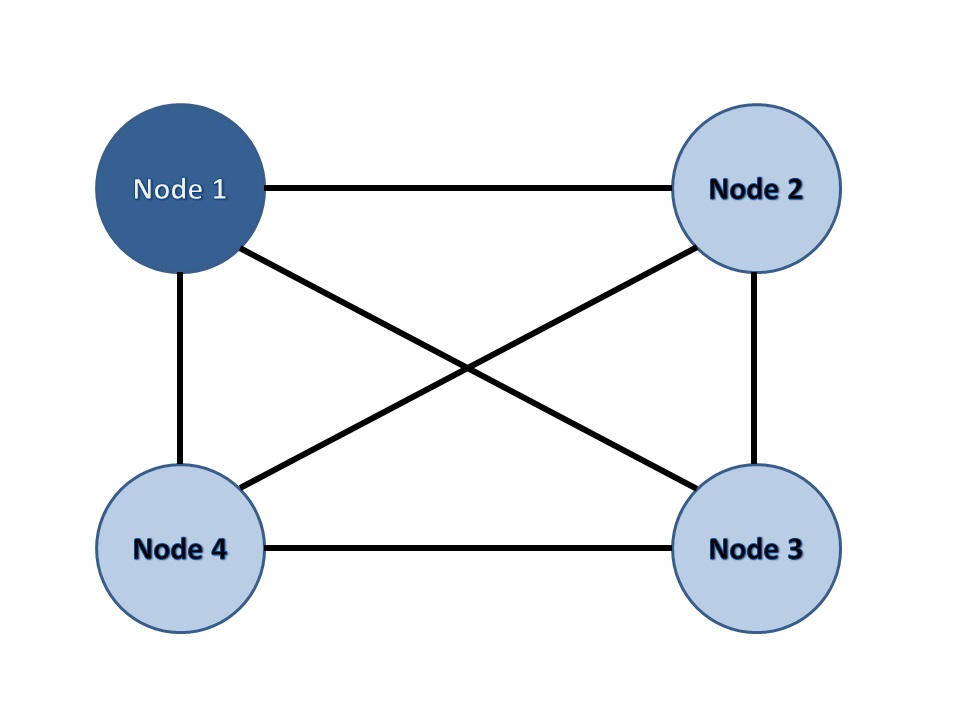}
 		    } 
 	       \subfigure[Network Structure]{
 	       	\includegraphics[width=0.23\textwidth]{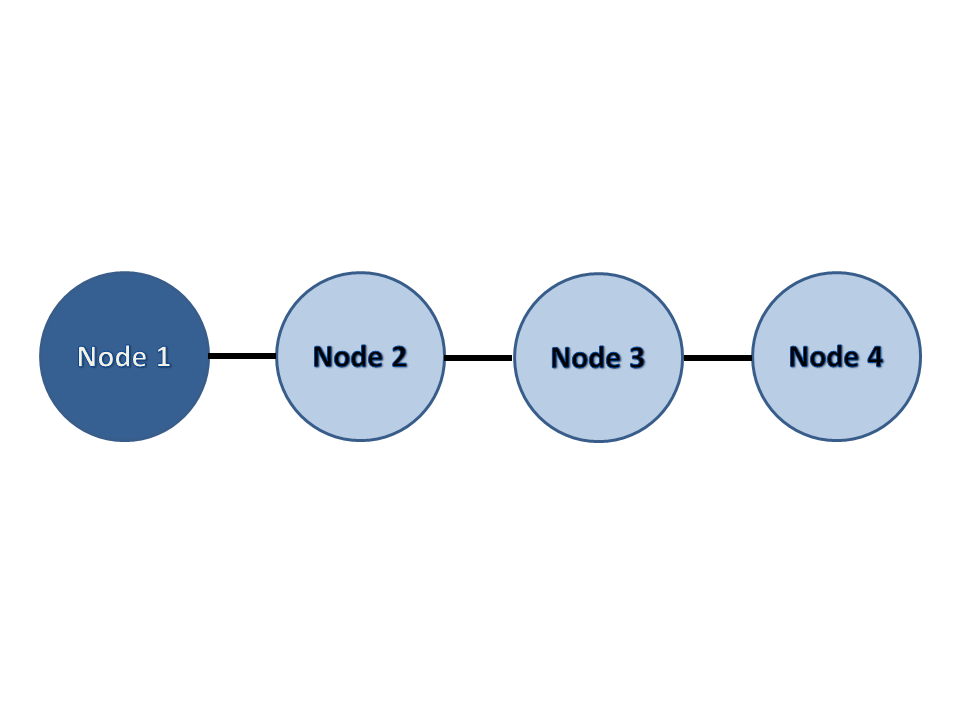}
 	       } \\
 		    \subfigure[Shock Diffusion Process]{
 		    	\includegraphics[width=0.24\textwidth]{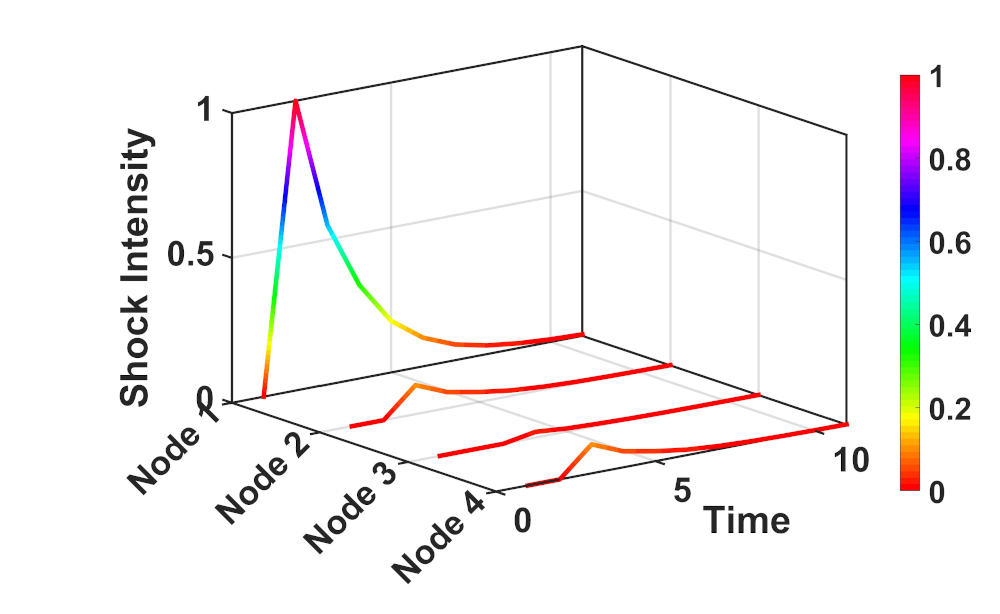}
 		    }
 			\subfigure[Shock Diffusion Process]{
 				\includegraphics[width=0.23\textwidth]{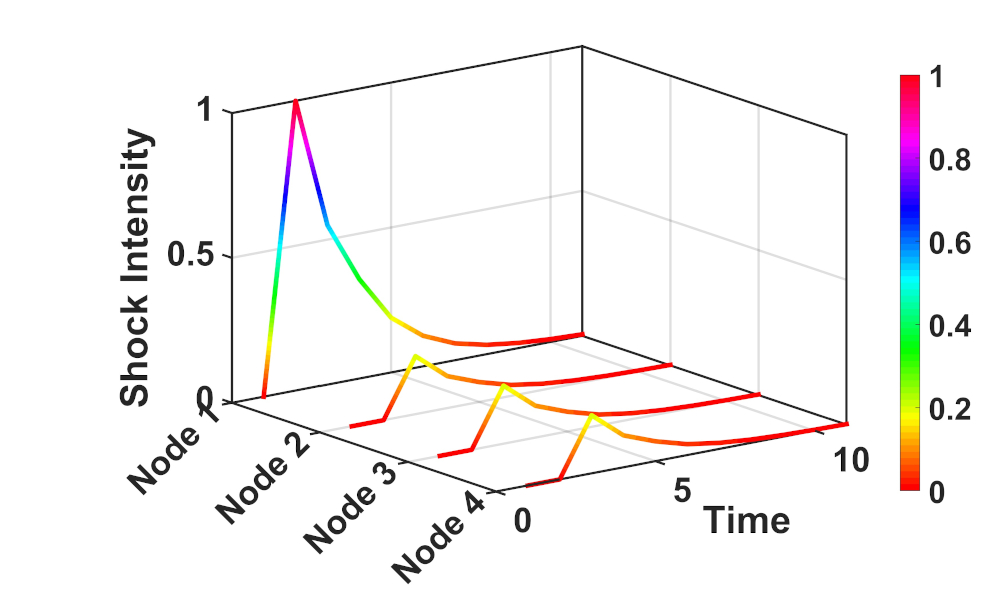}
 			}
 		  	\subfigure[Shock Diffusion Process]{
 		  	\includegraphics[width=0.23\textwidth]{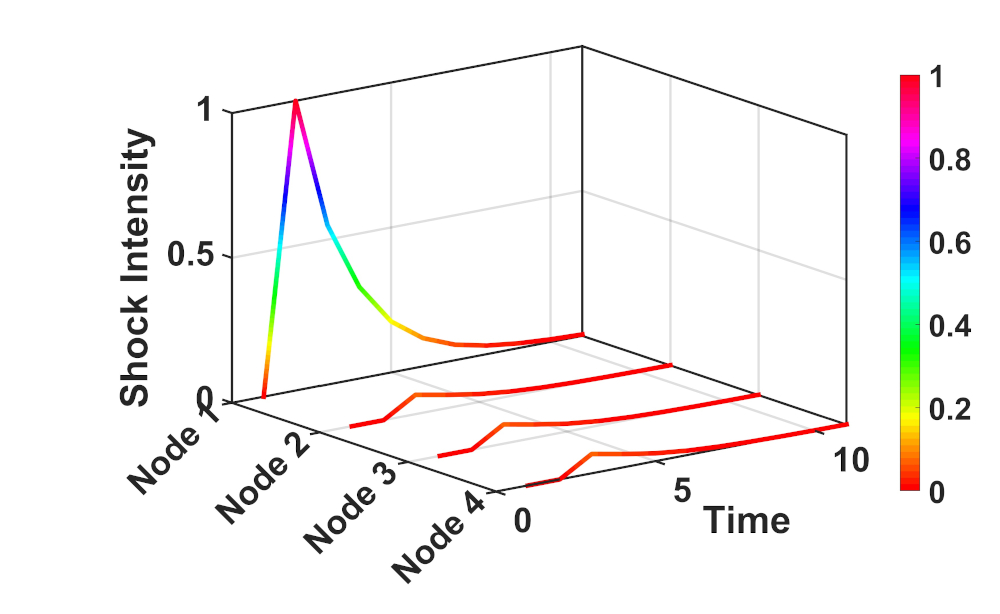}
 		  }
 	      \subfigure[Shock Diffusion Process]{
 	      	\includegraphics[width=0.24\textwidth]{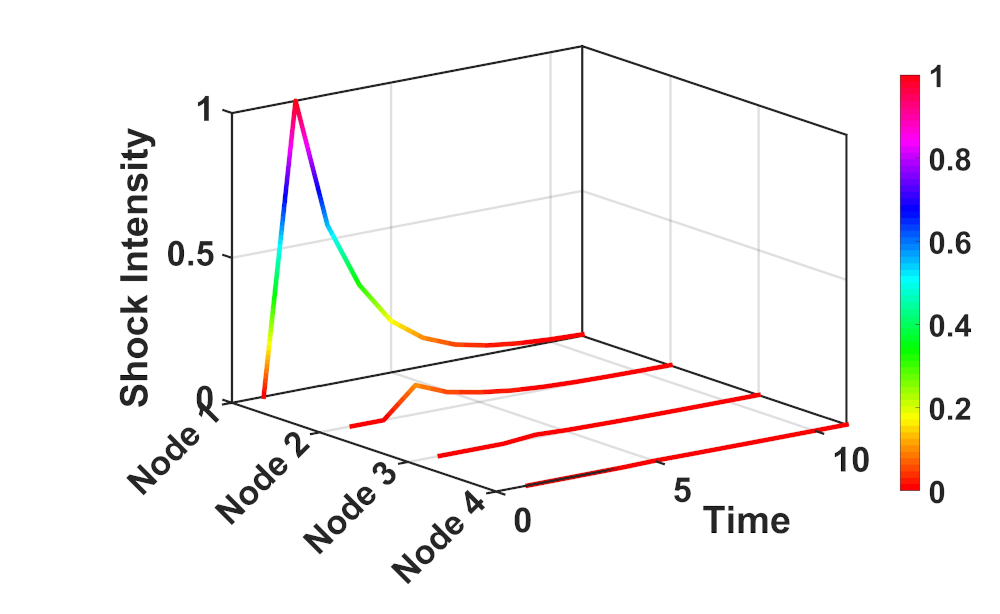}
 	      }
       \caption{\textbf{Distress propagation in toy models}. The plots show the usage of VAR(1) model to demonstrate the varying effect of different network structures on distress propagation through the network over time.}
         \label{fig: Toy_Models}
 \end{figure}

In Fig.~\ref{fig: Toy_Models}, we provide the toy network structures with the epicenters of distress (top panels; epicenter has darker shade) and the corresponding evolution of distress across the network (bottom panels). 
In panel (a), we have a circular network where each node is connected with its two immediate neighbors. Panel (b) shows a star-shaped core-periphery network. Panel (c) shows a complete graph. Panel (d) shows a linear graph. All networks have exactly four nodes for the sake of comparison.
In the bottom panels, we provide the impulse response functions corresponding to each node (nodes 1 to 4) over 10 time points. The intensity of the shock is captured by the height of the impulse response function (color coding shown by the side of the figures).

The nature of the shock propagation is quite intuitive. In panel (a), Node 1 is the epicenter which directly connects to Nodes 2 and 4. Both Nodes 2 and 4 are recipients of the shock but Node 3 which is in distance 2, does not react substantially (bottom panel). In panel (b), a shock to the core node (Node 1) equally impacts all the peripheral nodes.
For a complete graph (panel (c)), all nodes are symmetric and hence receives the shock in equal intensity. In panel (d), we have a linear graph and the epicenter is Node 1 which is the left-most node. As the impulse response functions in the bottom panels indicate, the shock propagates to other nodes but with lesser and lesser intensity. Node 4 (which is in distance 3 from the epicenter) shows almost no response.
The main take away from this simple exercise is that a core-periphery structure is the most conducive for distress propagation in an asymmetric manner (periphery to core shock propagation can be shown to be have mild impact and periphery to periphery through core would be almost negligible).

\subsubsection*{Distress propagation in empirical network of communities}

We implement the VAR model on the empirical network of communities (coarse-grained network structures). As we have described above, we calibrate the interaction matrix to the adjacency matrix of the observed network. Then we choose an epicenter and give a shock to it. The resultant impulse response functions allow us to study distress propagation over time and across the network.
First, we note that for each quarter there are two types of networks of communities; the first one is constructed by applying the modularity maximization algorithm, the second one is constructed by applying the Infomap algorithm. We have already noted before that the network of communities obtained from the Infomap algorithm is more coarse-grained than the one obtained from
modularity maximization algorithm (see Fig.~S2). Also, it is worthwhile to note from Table \ref{table:assortativity} that the network of communities obtained from the Infomap algorithm substantially less disassortative (very mildly assortative) than the one obtained from modularity maximization algorithm.

\begin{figure}[]
			\centering
			\subfigure[Network Structure]{
				\includegraphics[width=.41\textwidth]{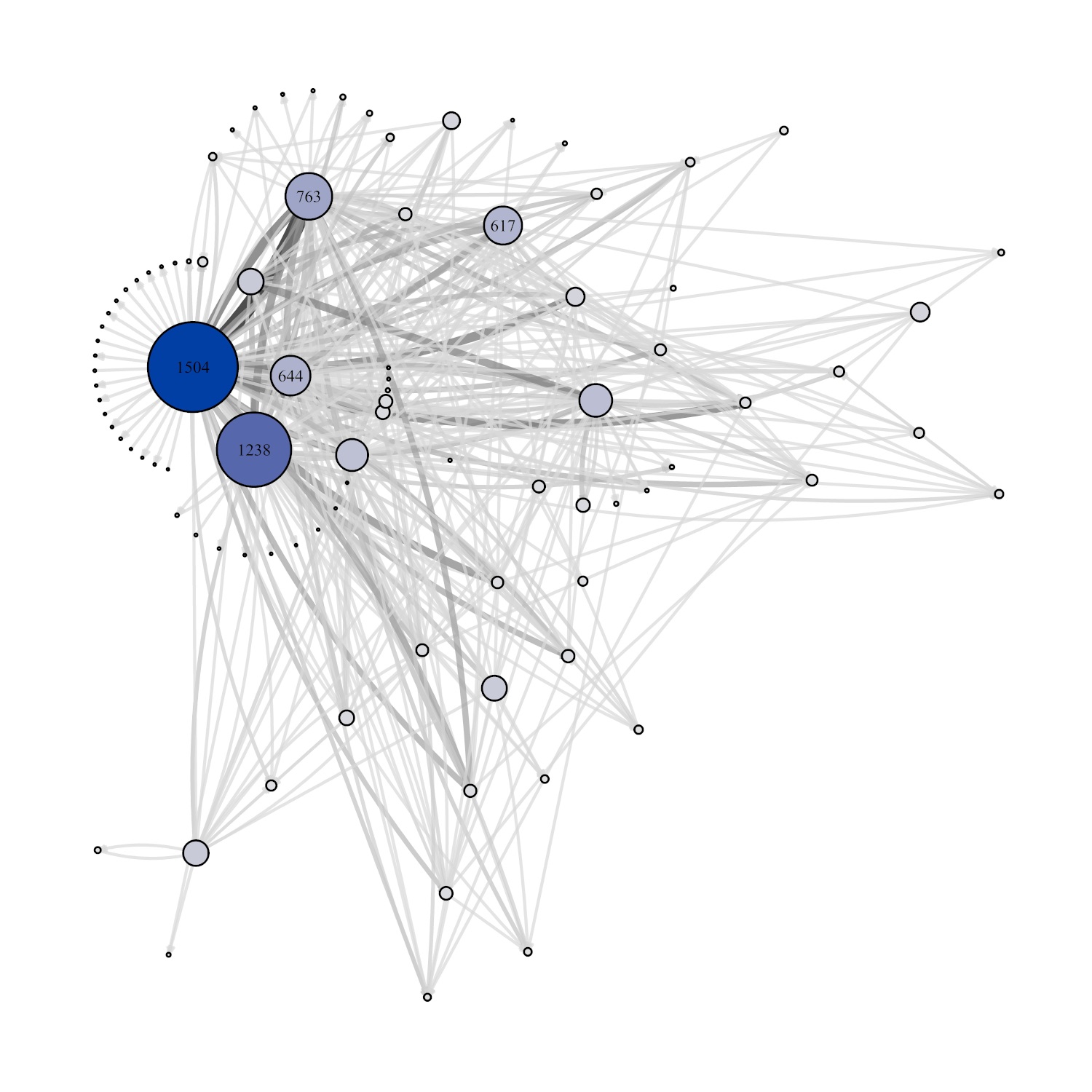}
			} 
			\subfigure[Shock Diffusion Process]{
				\includegraphics[width=.41\textwidth]{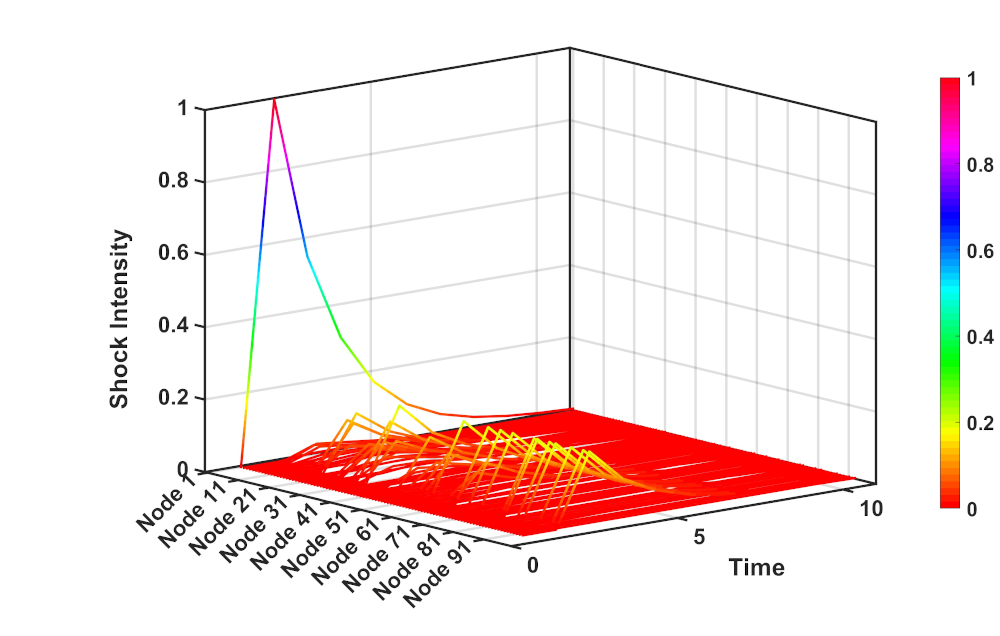}
			}
		\caption{\textbf{Distress propagation on the network of the communities (modularity maximization algorithm) from backbone network of quarter 2}:- Panel (a): Communities are depicted as nodes and their sizes are scaled to the number of constituent firms. We have labeled nodes (communities) with at least 600 firms. There are 5383 nodes in total. While constructing the network, we have removed self loops and nodes with degree zero. The resulting network consists of 93 nodes and 389 edges.
		Panel (b): Distress propagation through the network where the epicenter of the shock is the largest community (node with size 1504). Wide diffusion of shocks is evident from the impulse response functions.}
	\label{fig:modmax_largestcommunity_q2}
\end{figure}

\begin{figure}[]
			\centering
			\subfigure[Network Structure]{
				\includegraphics[width=.41\textwidth]{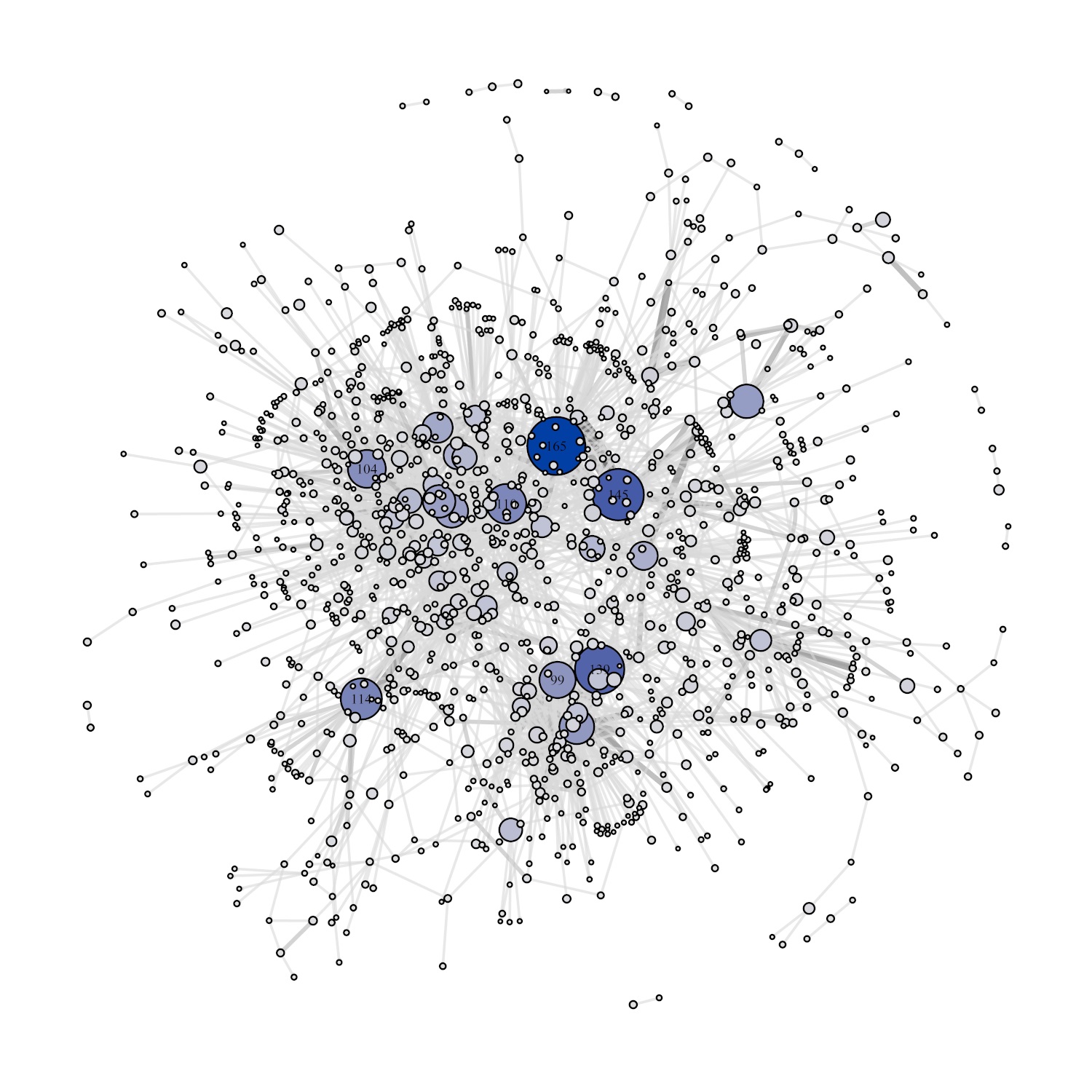}
			} 
			\subfigure[Shock Diffusion Process]{
				\includegraphics[width=.41\textwidth]{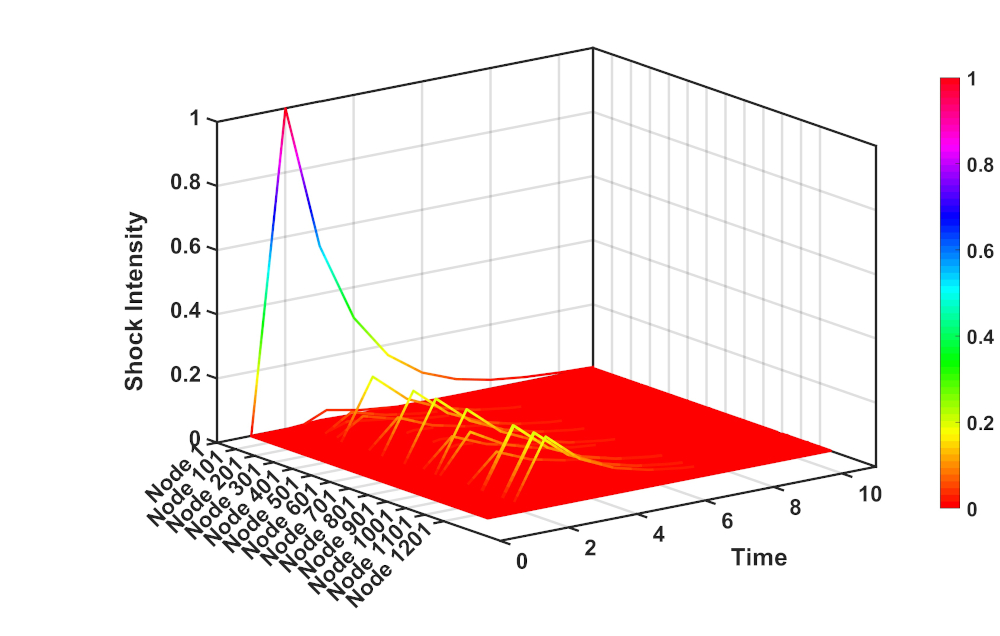}
			}
			\caption{\textbf{Distress propagation on the network of the communities (Infomap algorithm) from backbone network of quarter 2}: Panel (a): Communities are depicted as nodes and their sizes are scaled to the number of constituent firms. We have labeled nodes (communities) with at least 600 firms. There are 6540 nodes in total. While constructing the network, we have removed self loops and nodes with degree zero. The resulting network consists of 1257 nodes and 2538 edges.
				Panel (b): Distress propagation through the network where the epicenter of the shock is the largest community (node with size 166). Diffusion of shocks is evident from the impulse response functions. However, the impact is much more concentrated in a subset of nodes. In other words, distress propagation is more heterogeneous in terms of spread in the network, than in the case of modularity maximization algorithm.}
			\label{fig:infomap_largestcommunity_q2}
	\end{figure}

In Fig. \ref{fig:modmax_largestcommunity_q2}, we have plotted distress propagation on the network of communities constructed via modularity maximization algorithm in quarter 2 in the fiscal year 2016. Panel (a)
shows the network of communities. We give a unit shock to the largest node (denoting the largest community) and record the response across all nodes over time, as shown in panel (b). We see that a large number of nodes respond after one to three
periods and the peak response is about 20\% of the original shock (the $z$-axis denotes the shock intensity; colorbar shown in the figure).
We implement the same exercise on the same dataset where the network of communities has been constructed via Infomap algorithm. The results are shown in Fig. \ref{fig:infomap_largestcommunity_q2}. 
By nature of construction, modularity maximization recovers more granular network than Infomap consistently across all quarters. A notable difference from the earlier result (i.e. the network of communities based on modularity maximization) is that here 
the effects are much more heterogeneous; fewer number of nodes respond and the response intensity is also comparatively smaller in magnitude. Following an alternate approach, we studied distress diffusion on the networks of communities where the epicenter of the shocks are the nodes with highest eigenvector centrality.\cite{bardoscia2016distress,Acemoglu_Network_2012,banerjee2013diffusion,chakrabarti2018dispersion} 
We have provided complementary analysis of these experiments with all quarters (quarter 1-4, 2016 fiscal year) data in Figs.~S3-S4 (modularity maximization) and S5-S6 (Infomap) in the Supplementary information. Findings are both qualitatively and quantitatively similar. Shocks to the largest communities impact the system more than the community with highest centrality. This finding indicates that `too-big-to-fail' might be more dominant than `too-central-to-fail' entities.

Therefore, in this section we have established that the nature of coarse-grained modular structure of the network has important implications for the mechanism of shock propagation. A more disassortative network structure (obtained through modularity maximization; see Table \ref{table:assortativity}) leads to wider diffusion of shocks and the dynamic responses of the nodes are also larger in magnitude. We also see that   
`too-big-to-fail' modules seem to be more prominent for shock diffusion than `too-central-to-fail' modules in terms of impacting the network.

\section*{Summary and Discussion}

\noindent Distress propagation on complex networks is a dynamic process that is known to be affected by the topology of the
underlying networks.\cite{bardoscia2016distress} The recent literature in economics and finance recognizes the role
of the network architecture in the phenomena of economy-wide fluctuations.\cite{Acemoglu_Network_2012,acemoglu2015systemic,acemoglu2016networks,lux2009economics,starnini2019interconnected} However, one major problem is the
lack of granular data to quantify the network and analyze the mechanism of shock diffusion. In this paper, we leveraged on the
unique administrative data-set obtained from the Indian state of West Bengal that allowed us to construct
firm-to-firm buyer-seller network at the smallest level of granularity, by matching the tax records produced for each transaction by both parties of the transaction, i.e., the buying firm and the selling firm. Therefore, by matching the records we created a bilateral linkage between two firms and by collating all such linkages arising out of transactions within a given period of time (data had been collected every quarter in a fiscal year), we created the complete network of transactions across all the registered firms. This yielded networks of around 0.14 million firms in the database
with more than half million linkages, for every fiscal quarter of the year 2016.

Our main findings could be categorized under three heads: (i) The network showed large heterogeneity in degree connectivity, both in the first-order and second-order.\cite{Acemoglu_Network_2012}, (ii) The coarse-grained networks of communities,
created by applying modularity maximization algorithm exhibits moderate disassortativity with strong presence of multiple hub-and-spoke structures, whereas Infomap algorithm generates a modular structure with very mild assoratativity. (iii) The nature of distress propagation was substantially different across the networks of communities, viewed at different levels of coarseness. In particular, the more disassortative, coarser networks
gave rise to network-wide diffusion of shocks and the magnitude of the diffused shock was large; the more assortative, less granular network led to muted diffusion of shocks with a limited region of spread.
Our simulation results on the empirical coarse-grained structures indicate that `too-big-to-fail' nodes are more dominant than `too-central-to-fail' nodes in terms of distress propagation, which is a significant result in the context of identifying the relative contribution of size vis-a-vis centrality to network fragility.

Our results complement the current literature on diffusion processes on complex networks. Uniqueness of our data
provides an unprecedented view into the topological structure of production networks in the context of developing countries. Some prior studies had analyzed the data from developed countries like Japan.\cite{chakraborty2018hierarchical,fujiwara2010large,iino2015community}
We note that there is an important difference between the production structure in Japan (developed economy) vis-a-vis India (developing economy).
The first, Japanese one, is known to be dependent on just-in-time production process relying on very fast and efficient supply chain, whereas in case of India (representative of developing economies in terms of production process), the supply chains are much less efficient and therefore, firms depend on buffer stocks. The macroeconomic implication is that a disruption in such a network might possibly take more time to diffuse through the network and therefore, there would be lagged effects on the network.
This is consistent with the stickiness that we find in the network structures, before and after the demonetization shock. Our results indicate that at least at the macro-level, the production network did not change substantially.

Finally, we would like to mention that there are two limitations of our present study about the mechanisms of distress propagation: (i) The network we studied belongs to one state within a federal union of India with 28 states and 8 union territories. Thus the observed network is a small subset of a larger network of firms across states (or even countries, if one consider international trade). Therefore, shocks emanating from one state, in principle, could be transmitted to the network we have studied via trading partners within West Bengal. This is amenable in our framework, where instead of individual communities, we would need to give shocks to a set of firms (affected by exogenous shocks) and simulate the shock diffusion processes. However, such data is currently not available to us. Furthermore, the computation over a country-wide firm-firm network (even without considering international linkages) would be immensely costly.
(ii) We have not studied how such connections across firms had formed initially. In other words, we have taken the network structure as given\cite{atalay2011network,foerster2011sectoral} and studied the process of distress propagation on the network. In a complex world of evolving interconnections\cite{acemoglu2016innovation,arthur1999complexity,lux2009economics,starnini2019interconnected}, it would be interesting to find how the network evolves and accordingly how the shock diffusion process alters.\cite{newman2002spread,mones2014shock} Given the data limitations, presently it was not possible to conduct such an analysis.

\section*{Methods} 
\subsection*{Data description}
We have constructed the production network from quarterly tax data collected by the state officials of the state of West Bengal in India. The data had been collected under the Value-Added Tax (VAT) system introduced in India on 1st April, 2005. For the sake of completeness, we note here that this tax-collection scheme was substituted by Goods and Services Tax (GST) scheme on 1st July, 2017. However, our data-set does not overlap with the GST regime.  
In the tax collection scheme, each firm had an unique identification number. 
Due to the tax collection structure, both the buyer and the seller parties of a transaction were supposed to produce a record of the transaction amounts-- the nature of the transactions and amount of money exchanged, along with the identification number of the trading partner. Then records from both sides were matched to rule out false reports of trades. Therefore, a successful match between the buyer's record and a seller's record indicated successful completion of a trade between the buyer and the seller. In our database, a successful trade shows up as a linkage between two firms. The state officials collected and aggregated such data through tax filing in every quarter of the year. Therefore, one could construct a network arising out of all the reported binary linkages in a given quarter.
We have collected data that spans over four consecutive quarters, viz. quarters 1-4, for the year 2016; two quarters before and two quarters after the episode of ``demonetization''. Hence, all our analyses in this paper, are over these four snapshots of the production network.

\subsection*{Disparity filter and backbone construction}
We have  used the algorithm proposed by Serrano et al.~\cite{serrano2009extracting} in order to find the backbone of the weighted production network. The disparity filter algorithm is designed to extract the network backbone, by considering relevant edges present at all the scales in the system and exploiting the local heterogeneity as well as the local correlations among the weights. It may be noted that the disparity filter has a cut-off parameter $\alpha$, which has to be supplied as an input and determines the number of edges that are filtered from the original network. The choice of this cut-off parameter $\alpha$ is arbitrary, and may be determined by the scale of backbone network we would like to generate. One may have to actually tune this value, if we have sectoral information and other details  of the production firms. However, an advantage of this filtering technique, is that it preserves the cut-off of the degree distribution, the form of the weight distribution, and the clustering coefficient. 

\subsection*{Community detection}
A community may be defined as a collection of nodes that have a higher probability of connecting to one another within that collection than to nodes without\cite{barabasi2016network}.
In our case, the structure of communities provides a valid statistical way of defining ``clusters'' within which ``shocks'' affect everybody equally; we are interested in exploring how the shock spreads among other communities in the economy. There exist numerous community detection algorithms. Here, we use two community detection algorithms: (i) maximizing the modularity index\cite{newman2004fast} and (ii) Infomap (map equation) method\cite{rosvall2008maps}. 

Modularity gauges the magnitude of the partition of a network into smaller modules or communities. This method of partitioning is dependent on comparing the ratio of links in given communities to the expected number of links in a randomly rewired network preserving the degree distribution. In recent papers\cite{fujiwara2010large,iino2015community}, the modularity maximization technique had been used to study the community structure of the Japanese input-output network. This type of  community detection algorithm is based on the maximum modularity hypothesis, where the partition with the largest modularity is obtained. We have utilized the modularity maximization algorithm proposed by Newman\cite{newman2004fast}, which is essentially an iterative algorithm that connects pairs of communities in every step if the operation increases the modularity of the partition.
However, this class of algorithm suffers from a drawback that it does not detect very small communities, e.g, cliques, and has a limit of resolution\cite{fortunato2007resolution}. From our standpoint, this is a useful way of obtaining a ``meso-level'' view of the network, where we partition our network into comparatively bigger communities than what is feasible and then run a shock propagation model to see how the shock propagates in that network.

In order to check the robustness of our results, we have supplemented our results by another algorithm called the Infomap (map-equation) method to detect communities in our production network, and then compare the result of the shock propagation model.
The info-map method\cite{rosvall2008maps,lancichinetti2009community} is based on information theory and  
is measured by modeling a diffusion process
through the network with an exogenously specified partition. The measurement is based on 
the diffusion process across communities
as well as within communities.

\bibliographystyle{naturemag-doi}
\bibliography{main}

\section*{Acknowledgements}
We are thankful to the Directorate of Commercial Taxes of West Bengal, India for the permission to use their data. 
This research was supported in part by the International Centre for Theoretical Sciences (ICTS) during a visit for AC and AK participating in the summer research program on Dynamics of Complex Systems (Code: ICTS/Prog-DCS2019/07).
ASC acknowledges the R\&P grant, IIM Ahmedabad for partial support of this research. The authors have benefited from discussions with Sugata Marjit, Jyotsna Jalan, Sanjay Moorjani, Shekhar Tomar and Abinash Mishra.

\section*{Author contributions}
ASC and AC designed research. AK, AC and TN processed and analyzed the data. AK and AC prepared all the figures. All authors were involved in writing the paper.

%\section*{Additional information}
%\textbf{Supplementary Information} accompanies this paper.\\
%\noindent
%\textbf{Competing financial interests:} The authors declare no competing financial interests.

\setcounter{figure}{0}
\setcounter{table}{0}
\renewcommand{\thefigure}{S\arabic{figure}}
\renewcommand{\thetable}{S\arabic{table}}

\newpage
%%%%%%%%%%%%%%%%%%%%%
%\input{paper_suppl}
%%%%%%%%%%%%%%%%%%%%%
\section*{Supplementary Information}

\begin{figure}[h!]
	\centering
	\includegraphics[width=0.5\textwidth]{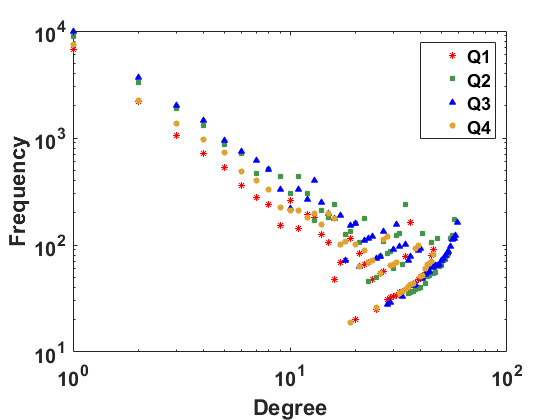}
	\caption{Second-order degree distributions of the backbone networks in four quarters.}
	\label{fig:Second_Degree_Dist}
\end{figure}

\begin{figure}[h!]
	\centering
	\includegraphics[width=0.3\linewidth]{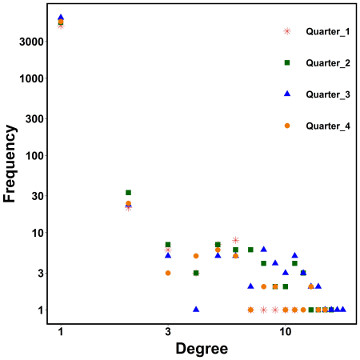}
	\llap{\parbox[b]{2.2in}{{\large\textsf{\textbf{a}}}\\\rule{0ex}{2in}}}
	\hskip 0.3in
	\includegraphics[width=0.3\linewidth]{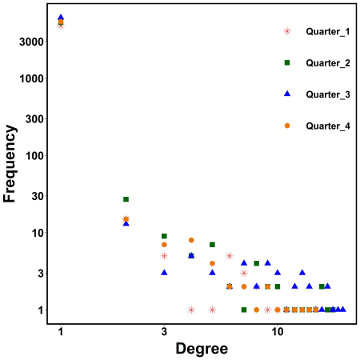}\llap{\parbox[b]{2.2in}{{\large\textsf{\textbf{b}}}\\\rule{0ex}{2in}}}\\
	\includegraphics[width=0.3\linewidth]{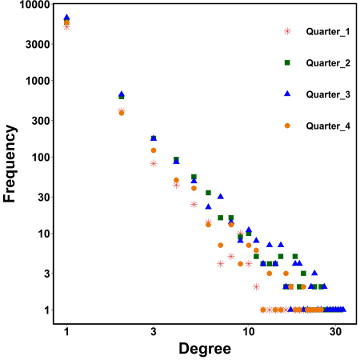}
	\llap{\parbox[b]{2.2in}{{\large\textsf{\textbf{c}}}\\\rule{0ex}{2in}}}
	\hskip 0.3in
	\includegraphics[width=0.3\linewidth]{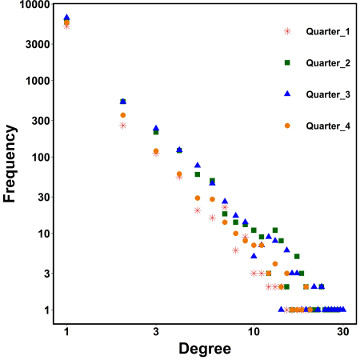}
	\llap{\parbox[b]{2.2in}{{\large\textsf{\textbf{d}}}\\\rule{0ex}{2in}}}	
	\caption{Degree distributions across four quarters: Panel (a) In-Degree with modularity maximization. Panel (b) Out-Degree with modularity maximization. Panel (c) In-Degree with Infomap. Panel (d) Out-Degree with Infomap.}
	\label{fig:Community_Degree}
\end{figure}

\begin{figure}[]
	\centering
	\subfigure[Quarter 1]{
		\includegraphics[width=.41\textwidth]{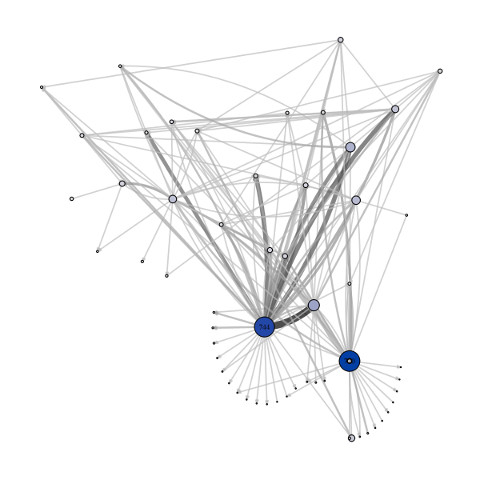}
	} 
	\subfigure[Quarter 2]{
		\includegraphics[width=.41\textwidth]{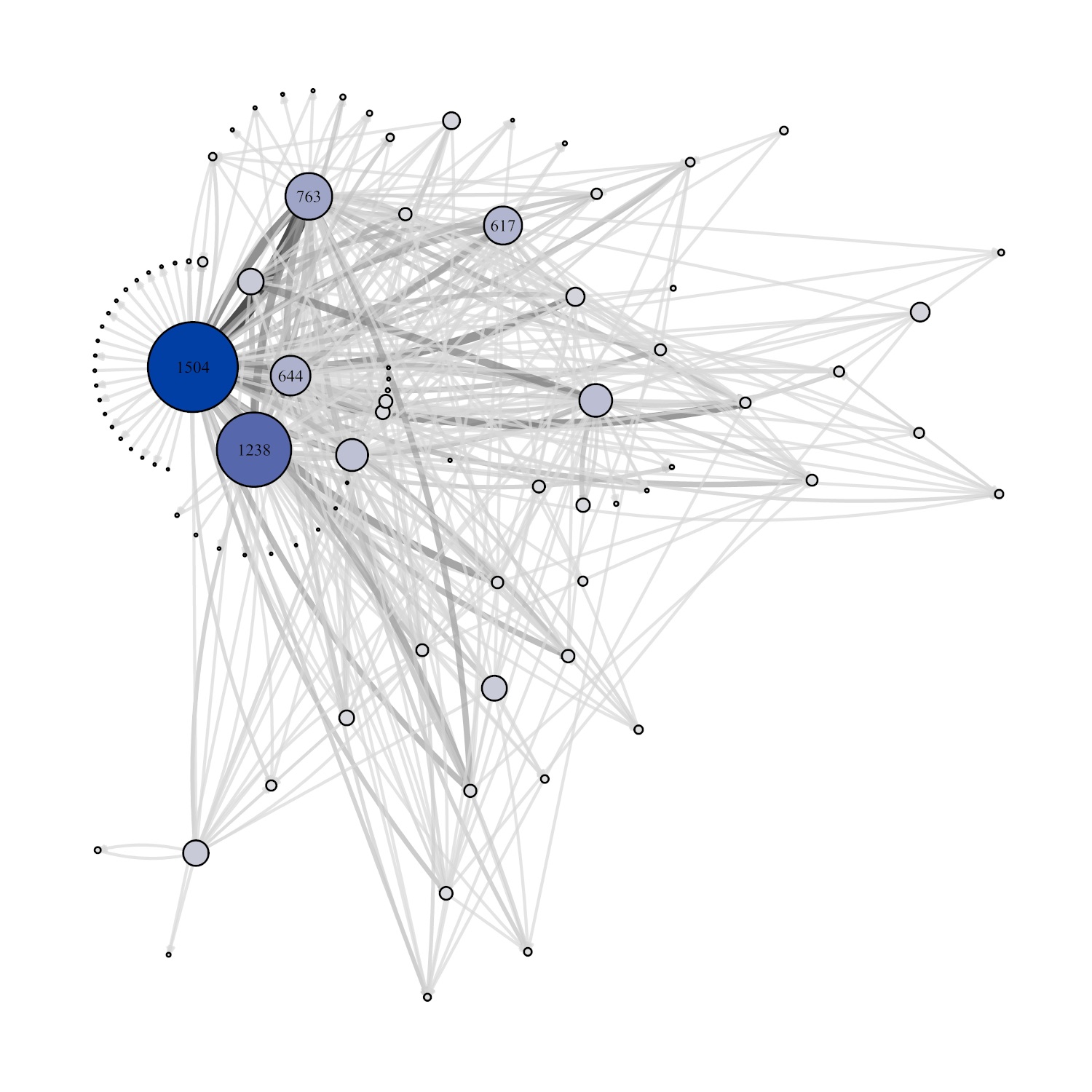}
	}
	\subfigure[Quarter 3]{
		\includegraphics[width=.41\textwidth]{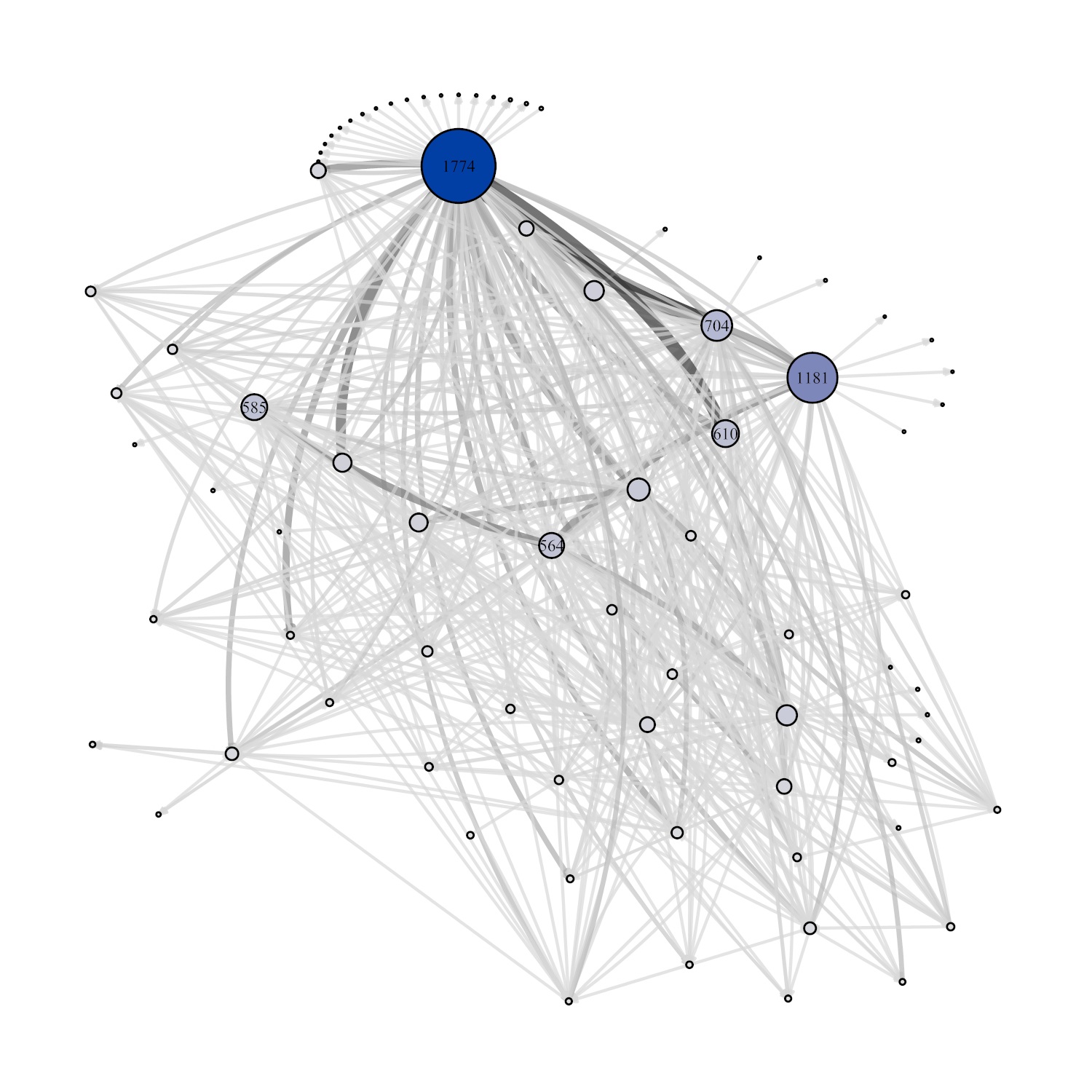}
	} 
	\subfigure[Quarter 4]{
		\includegraphics[width=.41\textwidth]{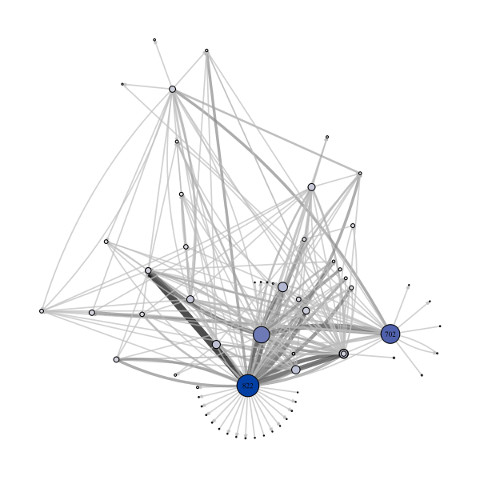}
	}
	\caption{\textbf{Networks of the communities (modularity maximization algorithm) from backbone networks for quarters 1-4}. Communities are depicted as nodes and their sizes are scaled to the number of constituent firms. We have labeled nodes (communities) with at least 600 firms. While constructing the network, we have removed self loops and nodes with degree zero. Panel (a): Q1- 58 nodes and 166 edges. Panel (b): Q2- 93 nodes and 389 edges. Panel (c): Q3- 79 nodes and 447 edges. Panel (d): Q4- 65 nodes and 225 edges.
	}
	\label{app:fig:modmax_q1234}			
\end{figure}

\begin{figure}[]
	\centering
	\subfigure[Q1: largest community]{
		\includegraphics[width=.41\textwidth]{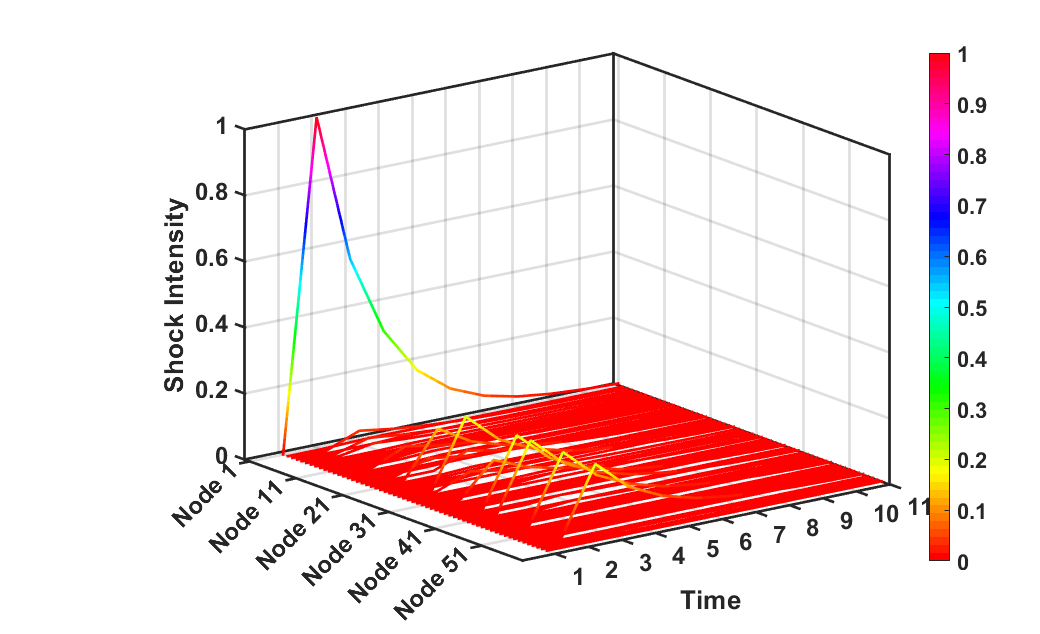}
	} 
	\subfigure[Q1: highest centrality]{
		\includegraphics[width=.41\textwidth]{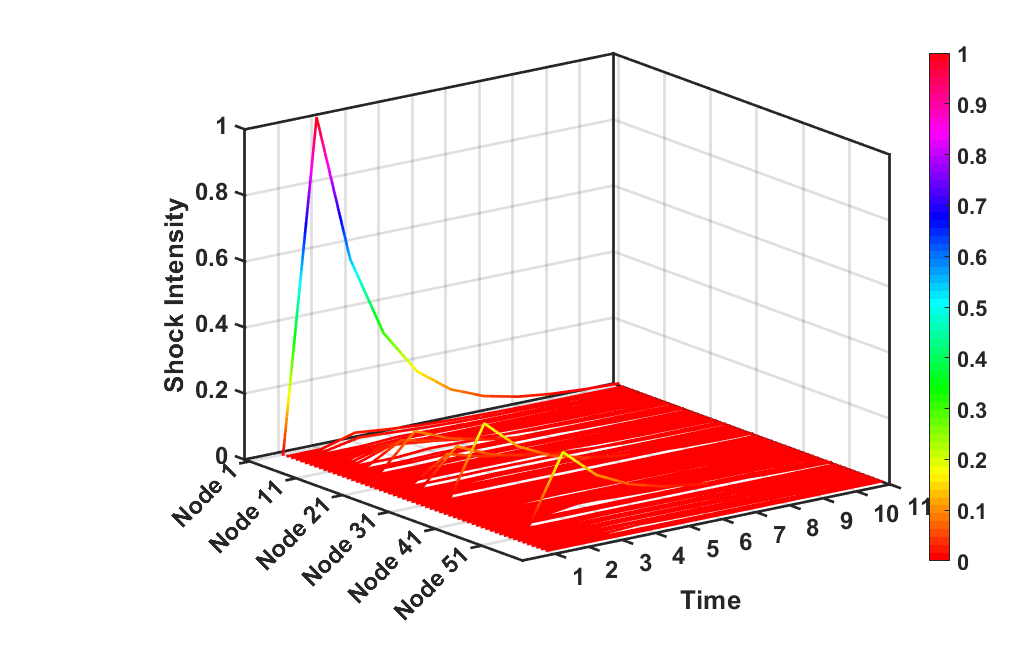}
	}\\
	\subfigure[Q2: largest community]{
		\includegraphics[width=.41\textwidth]{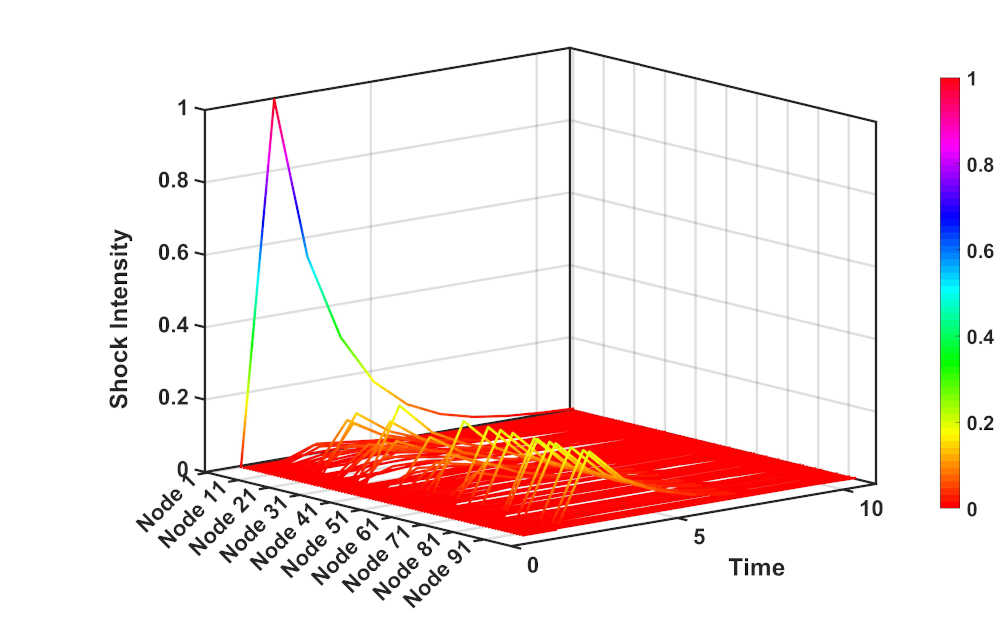}
	} 
	\subfigure[Q2: highest centrality]{
		\includegraphics[width=.41\textwidth]{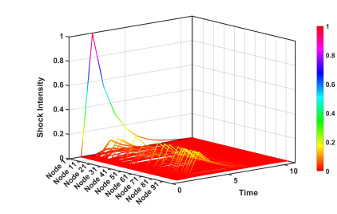}
	}\\
	\subfigure[Q3: largest community]{
		\includegraphics[width=.41\textwidth]{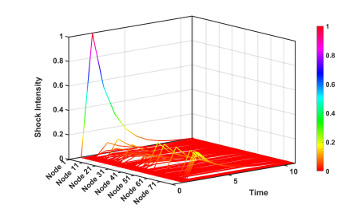}
	} 
	\subfigure[Q3: highest centrality]{
		\includegraphics[width=.41\textwidth]{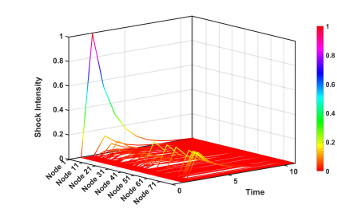}
	}\\
	\subfigure[Q4: largest community]{
		\includegraphics[width=.41\textwidth]{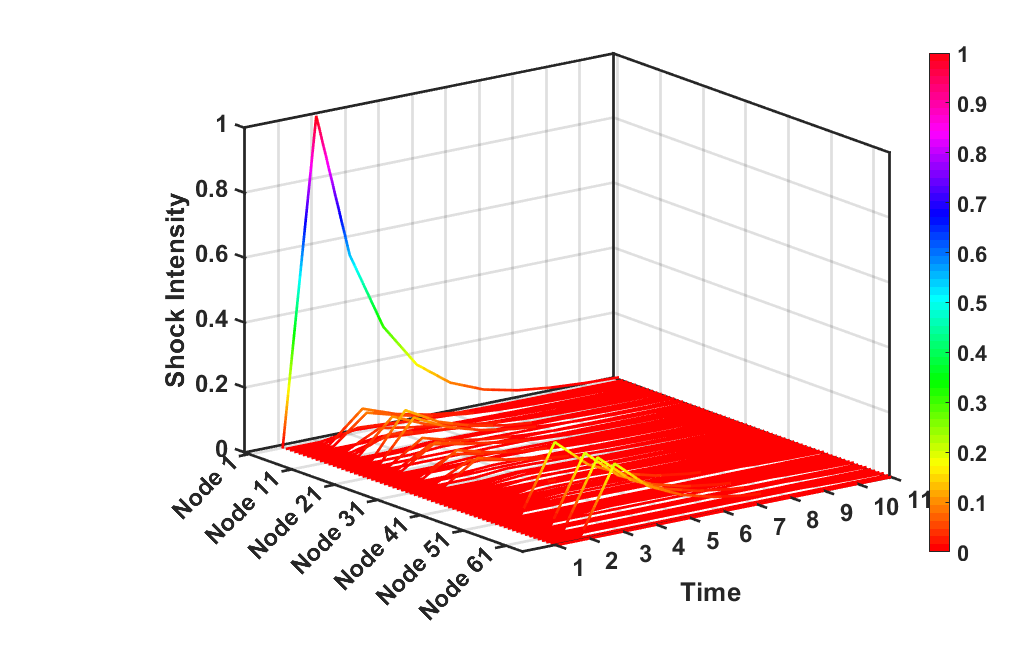}
	} 
	\subfigure[Q4: highest centrality]{
		\includegraphics[width=.41\textwidth]{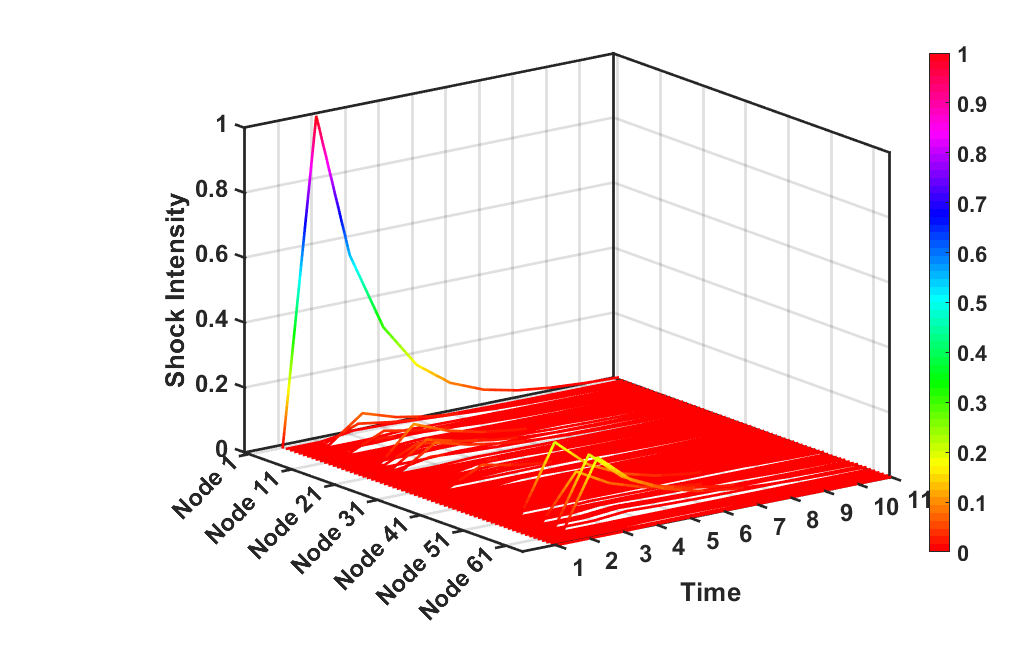}
	}
	\caption{\textbf{Impulse response functions corresponding to networks in Fig. \ref{app:fig:modmax_q1234} for quarters 1-4}. Panels (a, c, e, g): Epicenters are the largest communities in respective networks. Panels (b, d, f, h): Epicenters are the communities with largest centrality in respective networks. As evident, distress propagation initiated from the largest communities create more impact than distress initiated from the communities with highest centrality.
	}
	\label{app:fig:modmax_q1234_irf}			
\end{figure}

\begin{figure}[]
	\centering
	\subfigure[Quarter 1]{
		\includegraphics[width=.41\textwidth]{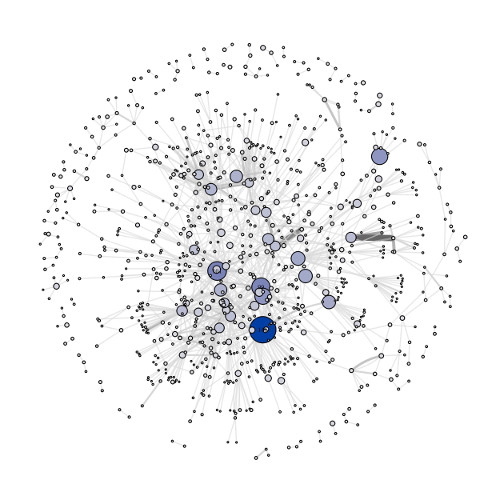}
	} 
	\subfigure[Quarter 2]{
		\includegraphics[width=.41\textwidth]{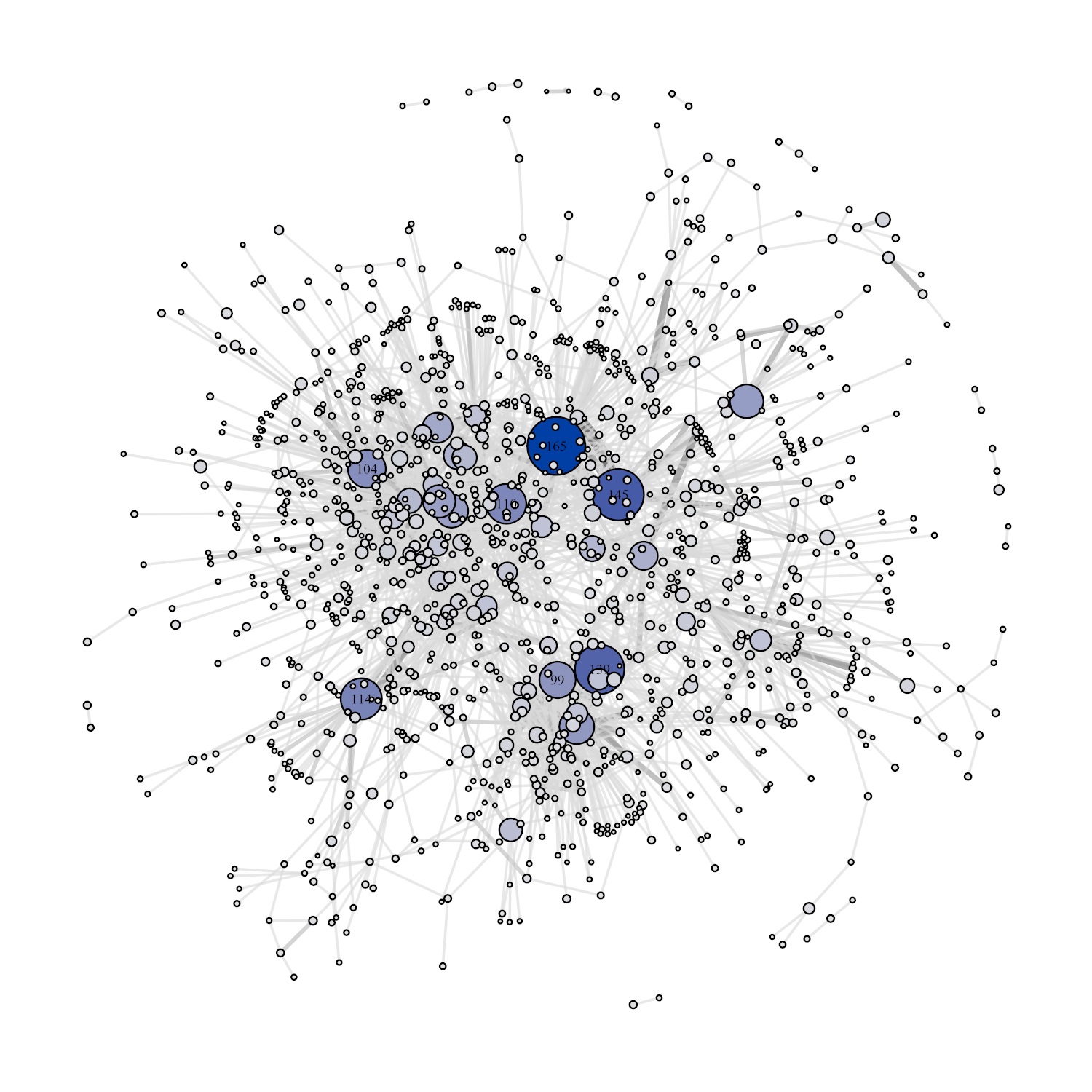}
	}
	\subfigure[Quarter 3]{
		\includegraphics[width=.41\textwidth]{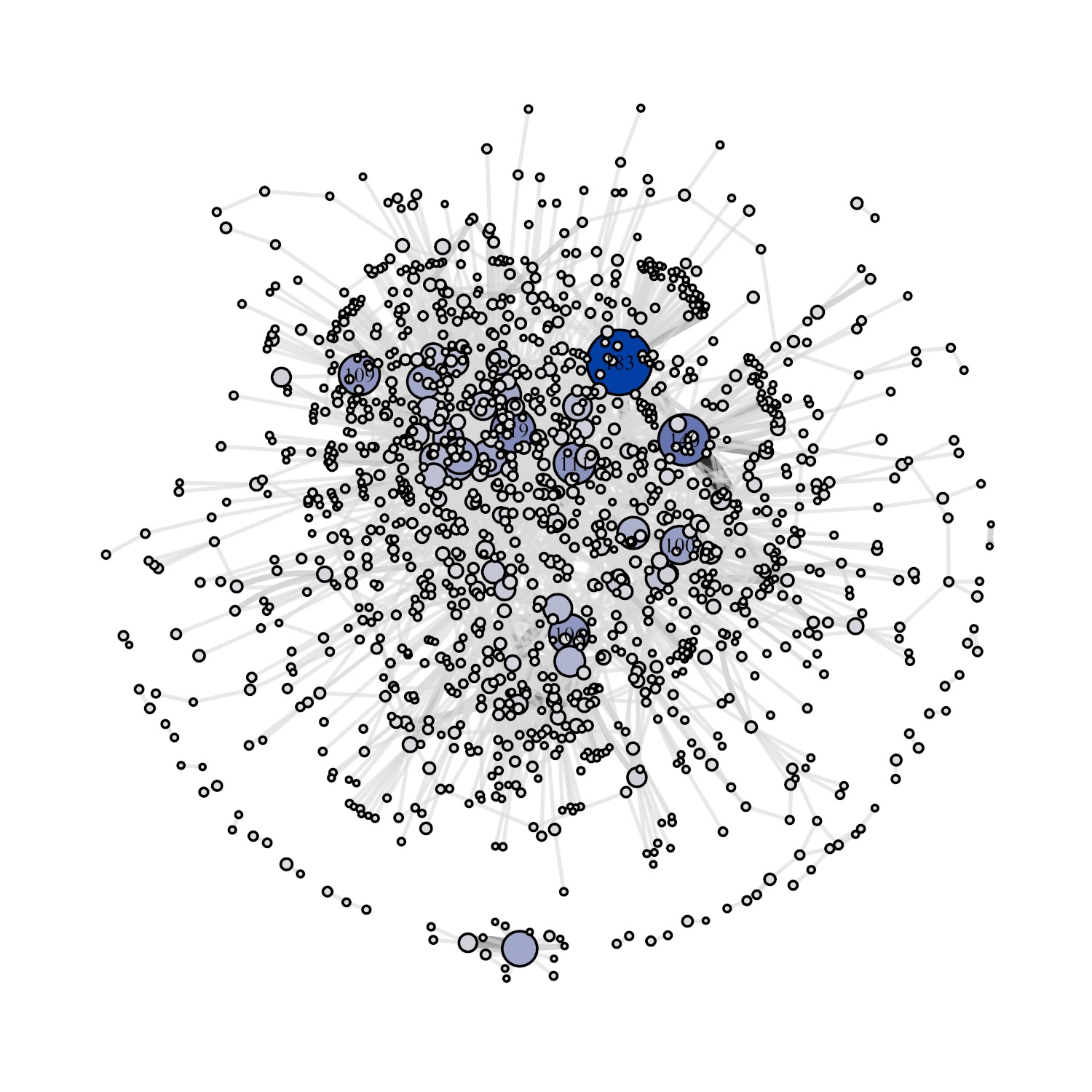}
	} 
	\subfigure[Quarter 4]{
		\includegraphics[width=.41\textwidth]{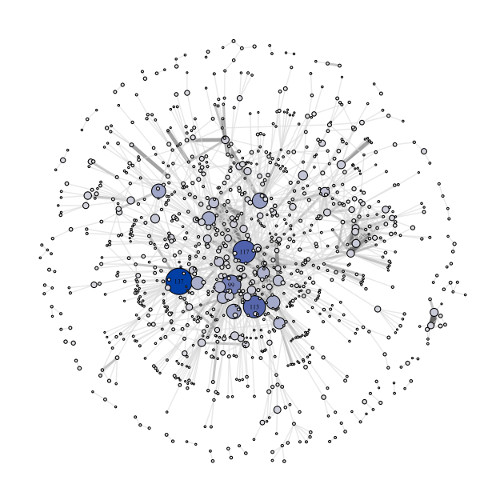}
	}
	\caption{\textbf{Networks of the communities (Infomap) from backbone networks for quarters 1-4}. Communities are depicted as nodes and their sizes are scaled to the number of constituent firms, after removing self-loops and nodes with degree 0. We have labeled nodes (communities) with at least 600 firms:- Panel (a): Q1- 909 nodes and 1316 edges. Panel (b): Q2- 1257 nodes and 2538 edges. Panel (c): Q3- 1343 nodes and 2718 edges. Panel (d): Q4- 1027 nodes and 1639 edges.
	}
	\label{app:fig:infomap_q1234}			
\end{figure}

\begin{figure}[]
	\centering
	\subfigure[Q1: largest community]{
		\includegraphics[width=.41\textwidth]{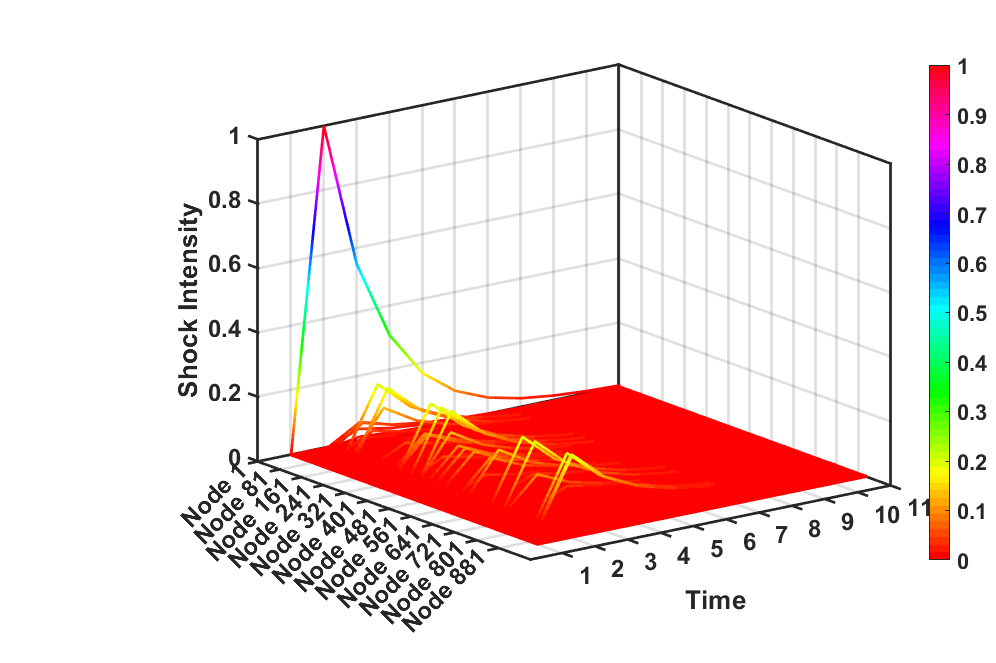}
	} 
	\subfigure[Q1: highest centrality]{
		\includegraphics[width=.41\textwidth]{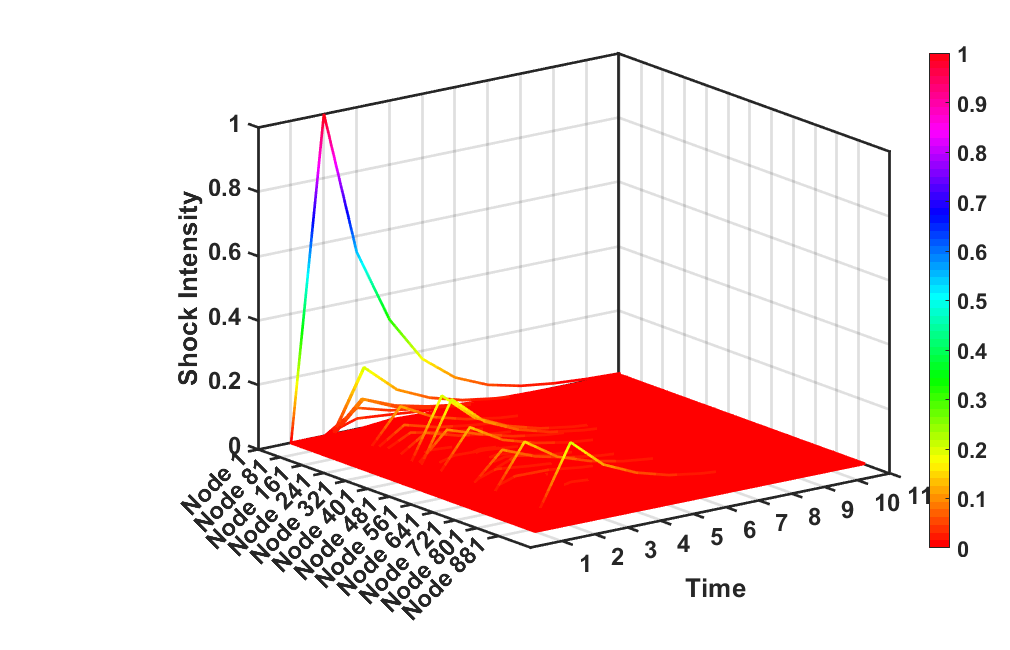}
	}
	\subfigure[Q2: largest community]{
		\includegraphics[width=.41\textwidth]{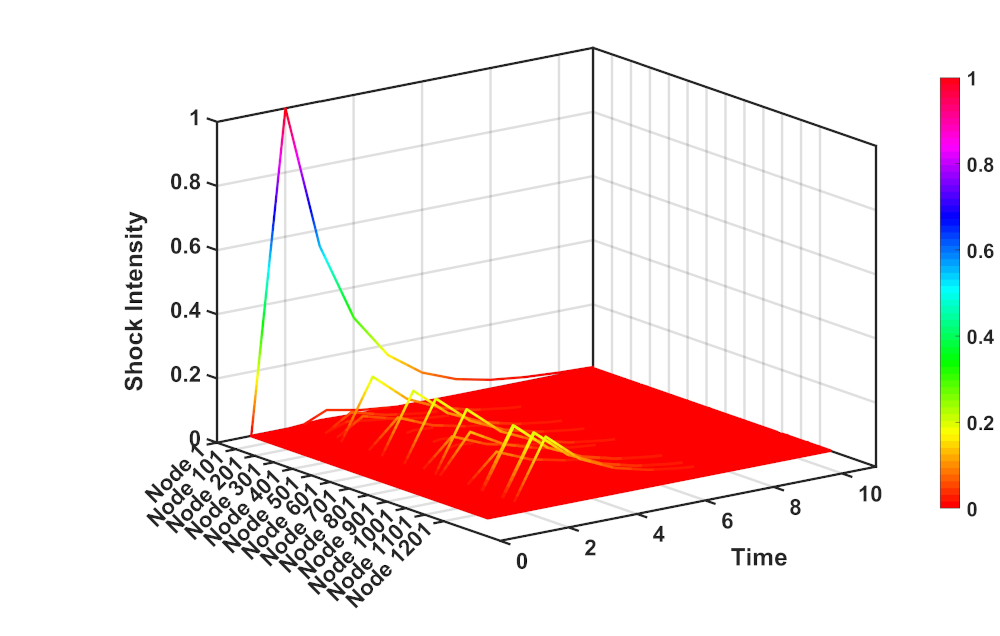}
	} 
	\subfigure[Q2: highest centrality]{
		\includegraphics[width=.41\textwidth]{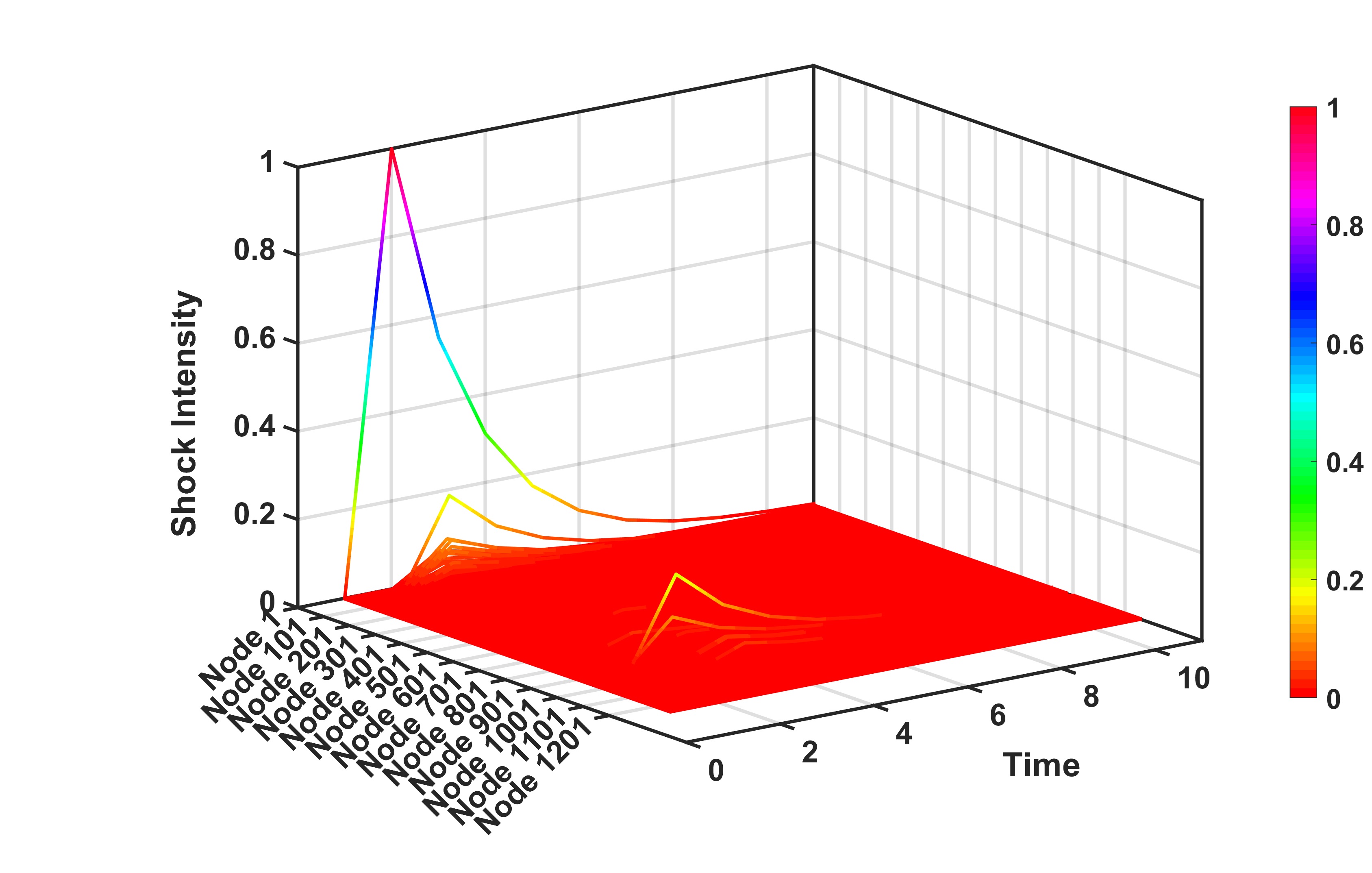}
	}
	\subfigure[Q3: largest community]{
		\includegraphics[width=.41\textwidth]{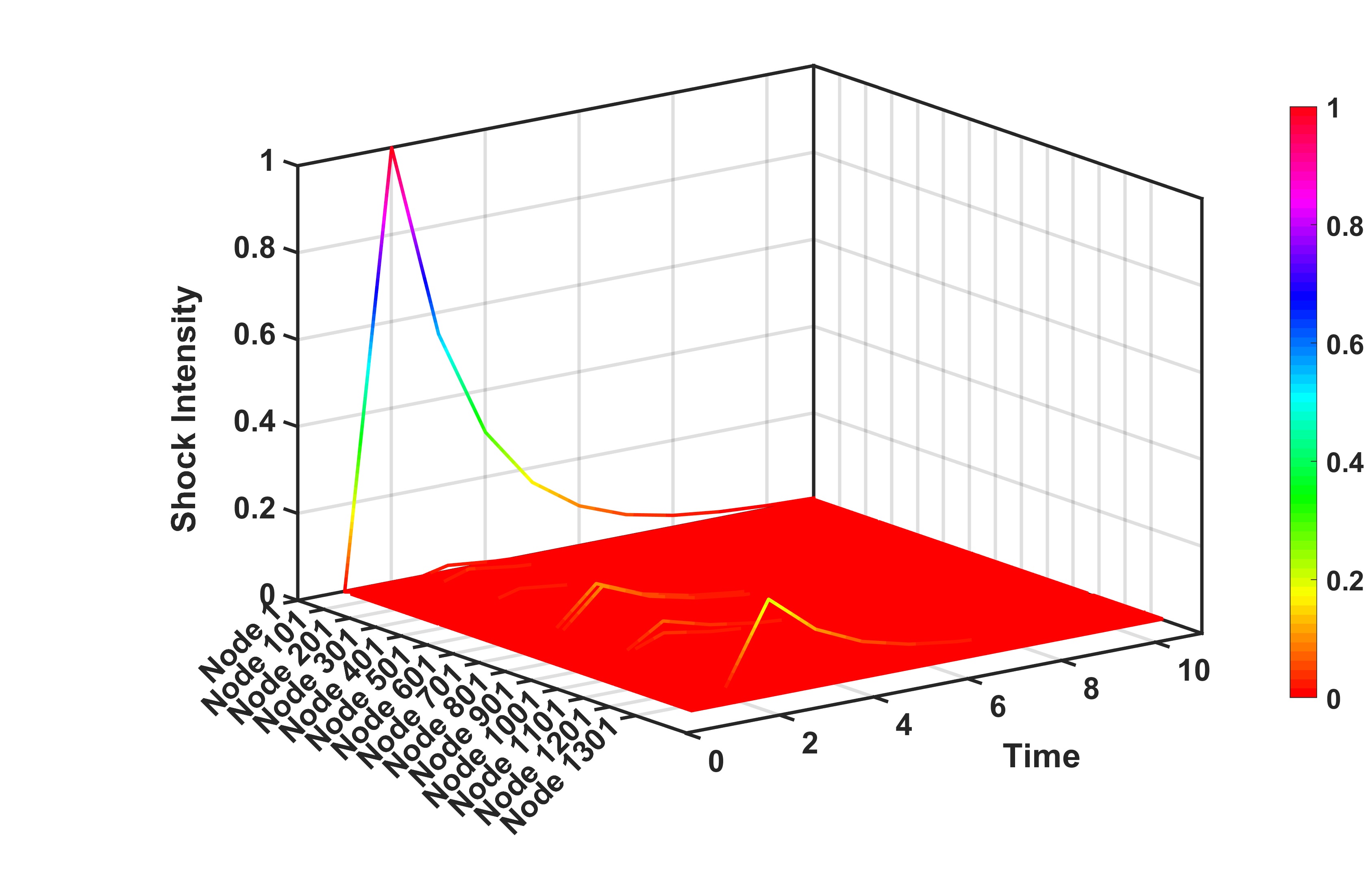}
	} 
	\subfigure[Q3: highest centrality]{
		\includegraphics[width=.41\textwidth]{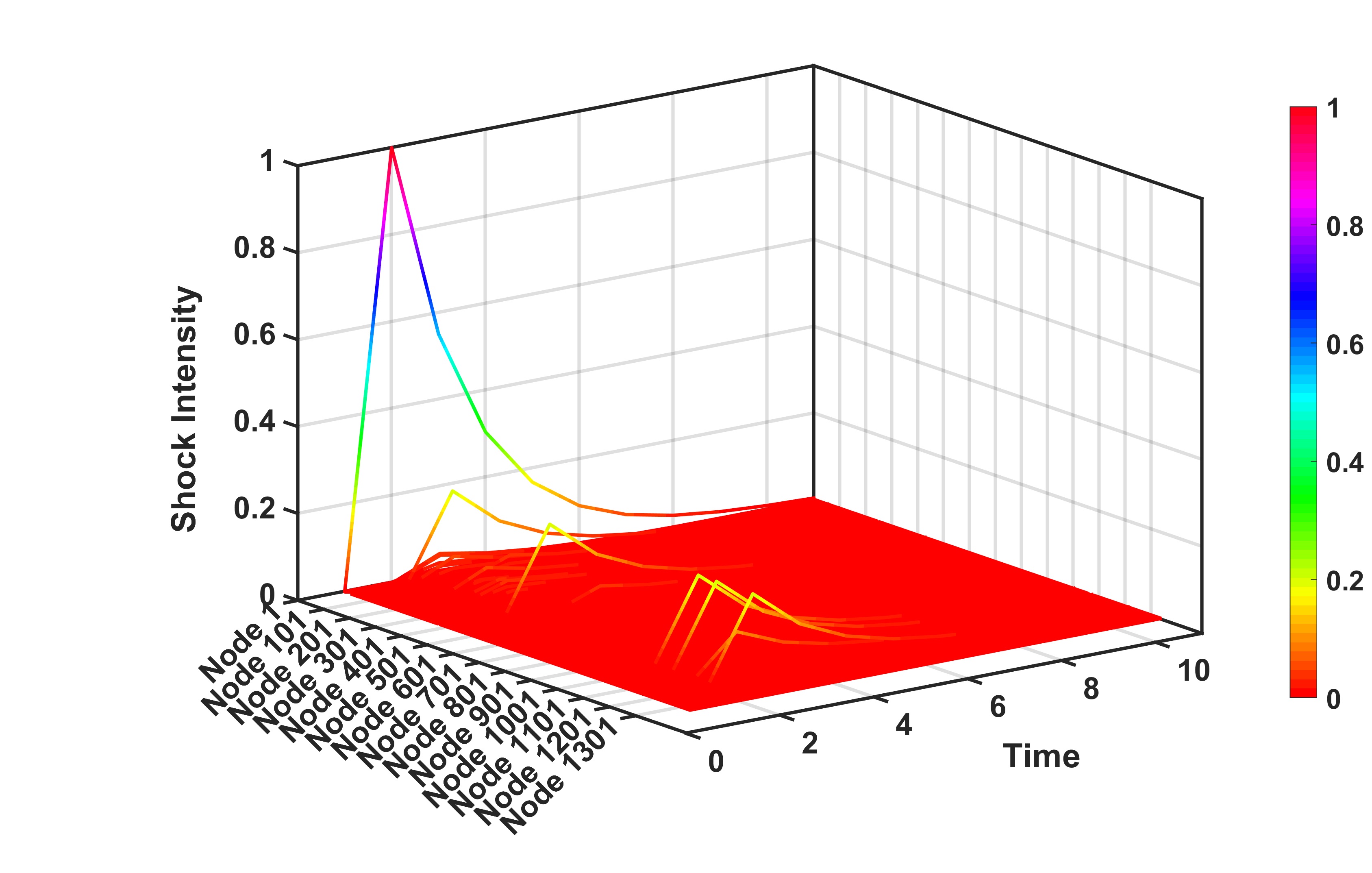}
	}
	\subfigure[Q4: largest community]{
		\includegraphics[width=.41\textwidth]{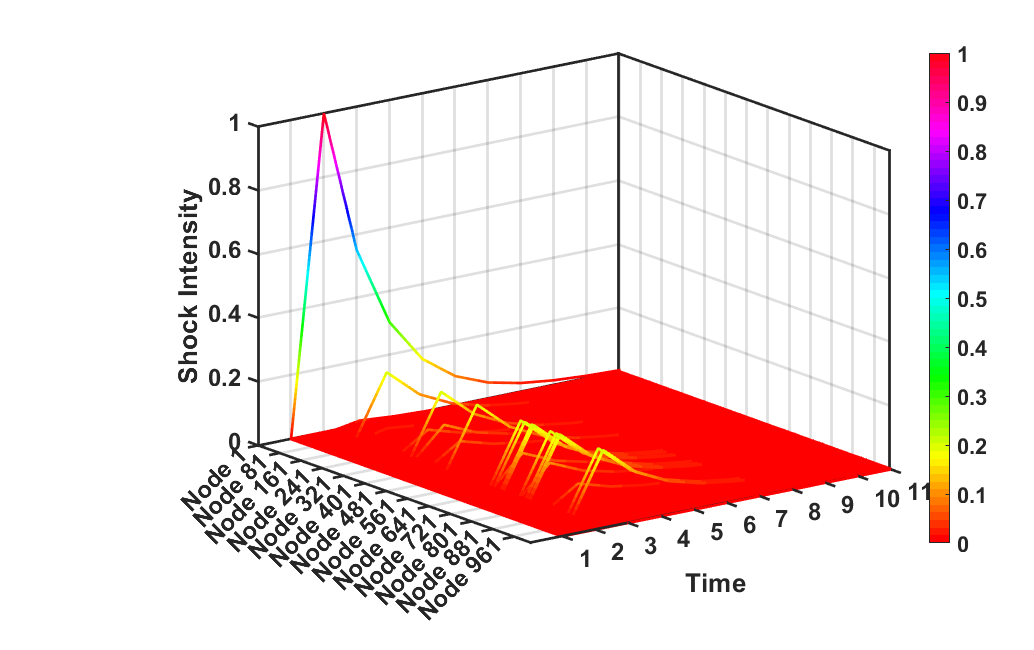}
	} 
	\subfigure[Q4: highest centrality]{
		\includegraphics[width=.41\textwidth]{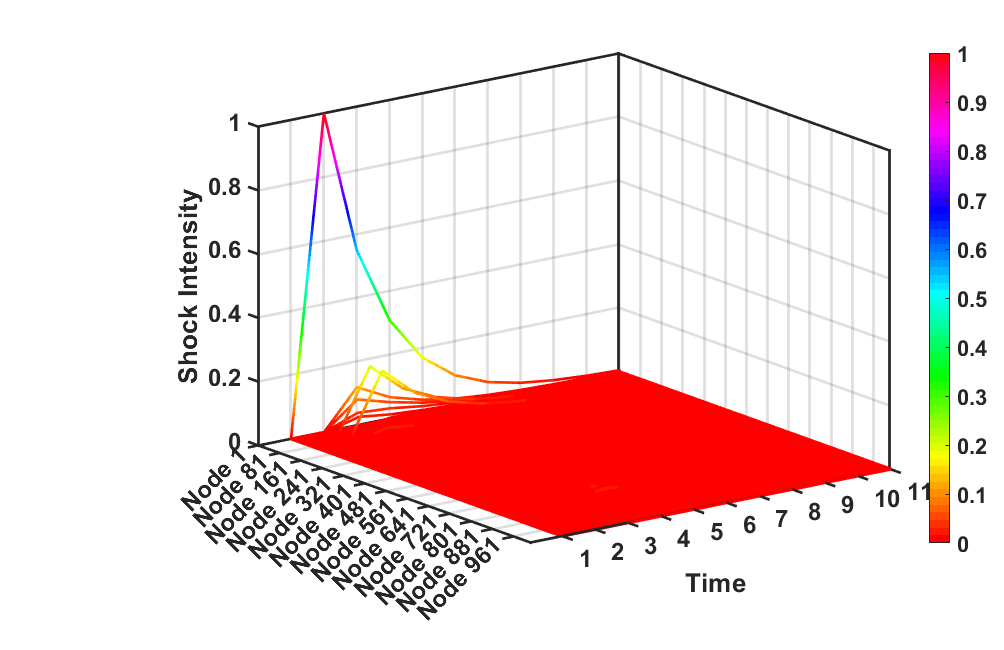}
	}
	\caption{\textbf{Impulse response functions corresponding to networks in Fig. \ref{app:fig:infomap_q1234} for quarters 1-4}. Panels (a, c, e, g): Epicenters are the largest communities in respective networks. Panels (b, d, f, h): Epicenters are the communities with largest centrality in respective networks. As evident, distress propagation initiated from the largest communities create more impact than distress initiated from the communities with highest centrality. In quarter 3, the spillover mechanism is muted compared to the rest of the quarters.
	}
	\label{app:fig:infomap_q1234_irf}			
\end{figure}

\end{document}